\documentclass[12pt]{aastex631}

\usepackage{color}
\usepackage{hyperref}
\usepackage{epstopdf}
\usepackage{graphicx}
\usepackage[FIGTOPCAP]{subfigure}
\usepackage{CJKutf8}
\usepackage{amsmath}
\usepackage{longtable}

\begin{document}
\title{A broken ``$\alpha$-intensity" relation caused by the  evolving  photosphere emission and the nature of the extraordinarily bright GRB~230307A}


\correspondingauthor{Yi-Zhong Fan}
\email{yzfan@pmo.ac.cn}

\author[0000-0002-8385-7848]{Yun Wang}
\affiliation{Key Laboratory of Dark Matter and Space Astronomy, Purple Mountain Observatory, Chinese Academy of Sciences, Nanjing 210034, China}

\author[0000-0003-4963-7275]{Zi-Qing Xia}
\affiliation{Key Laboratory of Dark Matter and Space Astronomy, Purple Mountain Observatory, Chinese Academy of Sciences, Nanjing 210034, China}
 
\author[0000-0001-6076-9522]{Tian-Ci Zheng}
\affiliation{Key Laboratory of Dark Matter and Space Astronomy, Purple Mountain Observatory, Chinese Academy of Sciences, Nanjing 210034, China}
\affiliation{School of Astronomy and Space Science, University of Science and Technology of China, Hefei, Anhui 230026, China}

\author[0000-0002-9037-8642]{Jia Ren}
\affiliation{School of Astronomy and Space Science, Nanjing University, Nanjing 210093, China}
\affiliation{Key Laboratory of Modern Astronomy and Astrophysics (Nanjing University), Ministry of Education, China}

\author[0000-0002-8966-6911]{Yi-Zhong Fan}
\affiliation{Key Laboratory of Dark Matter and Space Astronomy, Purple Mountain Observatory, Chinese Academy of Sciences, Nanjing 210034, China}
\affiliation{School of Astronomy and Space Science, University of Science and Technology of China, Hefei, Anhui 230026, China}

\begin{abstract}
GRB 230307A is one of the brightest gamma-ray bursts detected so far. With the excellent observation of GRB 230307A by Fermi-GBM, we can reveal the details of prompt emission evolution. As found in high-time-resolution spectral analysis, the early low-energy spectral indices ($\alpha$) of this burst exceed the limit of synchrotron radiation ($\alpha=-2/3$), and gradually decreases with the energy flux ($F$). A tight $E_{\rm p}\propto F^{0.54}$ correlation anyhow holds within the whole duration of the burst, where $E_{\rm p}$ is the spectral peak energy. Such evolution pattern of $\alpha$ and $E_{\rm p}$ with intensity is called ``double tracking". For the $\alpha-F$ relation, we find a log Bayes factor $\sim$ 210 in favor of a smoothly broken power-law function over a linear function in log-linear space. We call this particular $\alpha-F$ relation as broken ``$\alpha$-intensity", and interpret it as the evolution of the ratio of thermal and non-thermal components, which is also the evolution of the photosphere. GRB 230307A with a duration of $\sim 35~\rm s$, if indeed at a redshift of $z=0.065$, is likely a neutron star merger event (i.e., it is intrinsically ``short"). Intriguingly, different from GRB 060614 and GRB 211211A, this long event is not composed of a hard spike followed by a soft tail, suggesting that the properties of the prompt emission light curves are not a good tracer of the astrophysical origins of the bursts. The other possibility of $z=3.87$ would point toward very peculiar nature of both GRB 230307A and its late time thermal-like emission.



\end{abstract}

\section{Introduction} \label{sec:intro}
At the early time when gamma-ray burst (GRB) were recognized as the origin of cosmology, the prompt emission was predicted to be quasi-thermal \citep{paczynski1986gamma,goodman1986gamma}.
This is due to the huge energy ($\gtrsim$ 10$^{53}$ erg) released in a very small space ($r \sim 10^7 - 10^8$ cm), which is bound to produce a huge optical depth.
However, in the BATSE era, most of the GRB spectra are non-thermal, which can be well fitted by the Band function \citep{band1993batse}.
The main theory is to explain the non-thermal component as synchrotron radiation and synchrotron self-Compton scattering from relativistic electrons \citep{tavani1996shock,lloyd2000synchrotron,zhang2002analysis,daigne2011reconciling,zhang2010internal,uhm2014fast}.
Although the synchrotron radiation model has been well developed in GRBs, there are still some problems. For example, the synchrotron radiation or synchrotron self-Compton scattering cannot explain the steep low-energy spectral index in some observations \citep{crider1997evolution,preece1998synchrotron,preece2002consistency,ghirlanda2003extremely}, and cannot explain the correlation between the peak energy and luminosity ($E_{\rm p}$ - $L$) without introducing additional assumptions \citep{golenetskii1983correlation,amati2002intrinsic}, and the efficiency of energy dissipation in the internal shock model \citep{mochkovitch1995internal,kobayashi1997can,panaitescu1999power,lazzati1999constraints,kumar1999gamma,spada2000analysis,guetta2001efficiency,maxham2009modeling}. Hybrid models of thermal and non-thermal components can address these issues \citep{pe2017photospheric}.

So far, the most direct evidence of photosphere emission is the observation of GRB 090902B by Fermi satellite \citep{abdo2009fermi}.
\cite{ryde2010identification} confirmed that the spectrum of GRB~090902B is a broadened Planck function superimposed with a power-law component extending to the high energy band, such geometric broadening \citep{pe2008temporal,lundman2013theory,deng2014low} can be described as a multi-color blackbody model \citep{ryde2010identification,hou2018multicolor}.
Events like GRB~090902 are very rare due to the identification of thermal components usually requires sufficient photon counts for time-resolved spectrum analysis.
The currently detected GRB~230307A, which may be the second brightest GRB \citep{GCN33414}, has a high-quality observation through the Fermi Gamma-ray Burst Monitor (GBM).
It thus provides a valuable opportunity to study the time-resolved spectrum of GRB.
It has been reported a Bad time interval of GBM (TTE: $T_0
+[3.00,~7.00]~\rm s$) due to the pulse pile-up~\citep{2023GCN.33551....1D}. Following their recommendation, we only take into account the data in the time intervals of $T_0+ [-0.02,~2.67]~\rm s$ and $T_0+ [7.22,80.14]~\rm s$ without packet losses.

The paper is organized as follows:
In Section \ref{sec:obs_ana}, we present the data analysis of GRB~230307A by the self-developed {\tt HEtools} package.
In Section \ref{sec:2}, we further characterize GRB~230307A based on the Bayesian inference results for time-resolved spectra.
In Section \ref{sec:3}, we discuss some of these results and compare this burst with some typical GRBs. The nature of GRB 230307A is also briefly examined.
In Section \ref{sec:4}, we summarize results of our analysis.

\section{Observation and Data analysis}\label{sec:obs_ana}
Shortly after the last monster swept by, Fermi-GBM reported another bright GRB~230307A \citep[trigger 699896651/ 230307656;][]{GCN33405}.
From the current observations and some preliminary analysis \citep{GCN33406,GCN33411,GCN33414,GCN33418,GCN33427}, its energy flux is second only to GRB~221009A.

\begin{figure}[!h]
    \centering
    \includegraphics[width=0.8\textwidth]{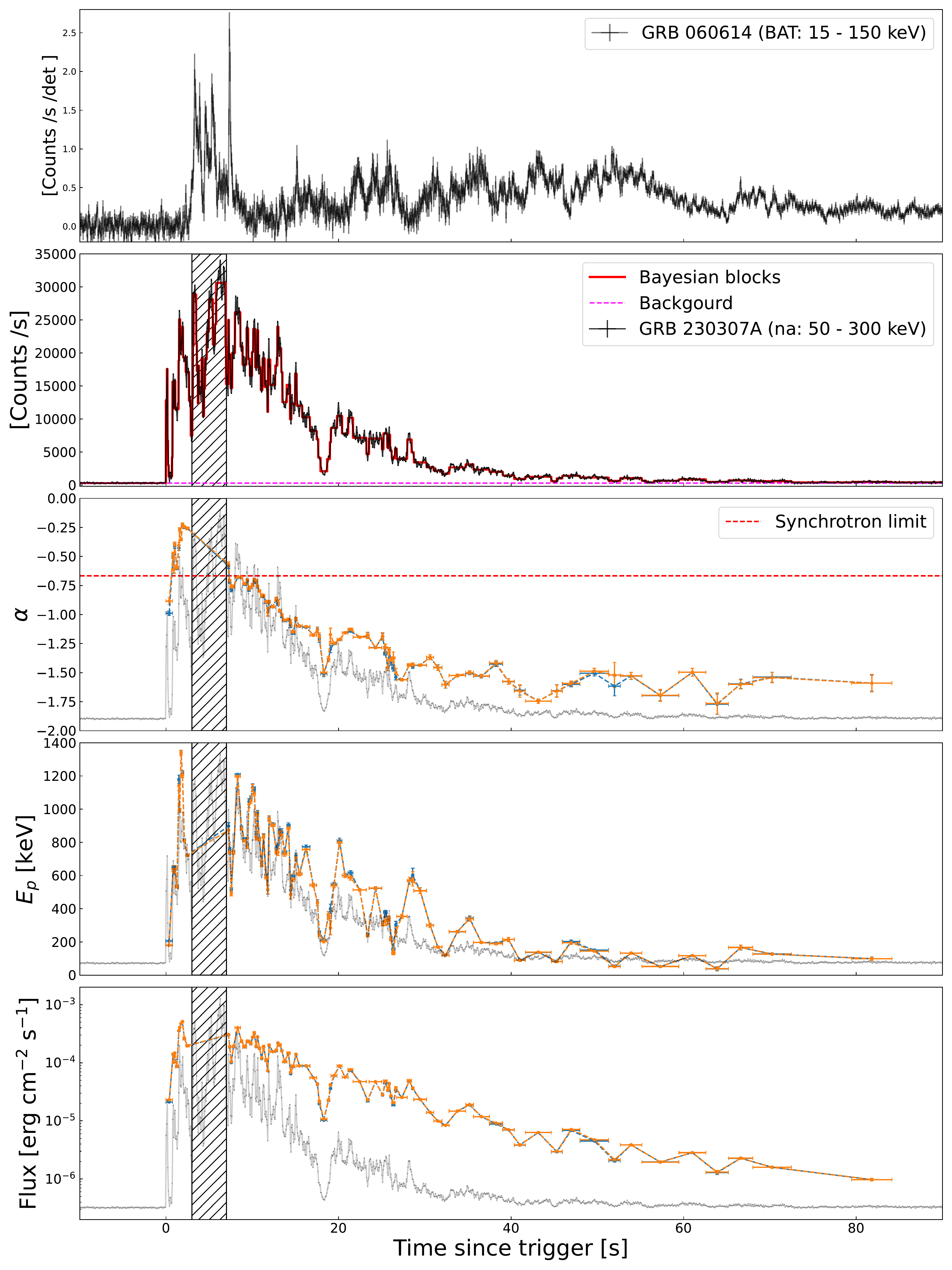}
    \caption{Observational data and parameter evolution of GRB~220307A.
    {{The first panel shows the light curve of GRB 060614, the well-known long-short burst (i.e., a long duration GRB from a compact object merger).}
    The second panel shows the light curves of GRB~220307A and the Bayesian block in the black and red solid line, respectively.} 
    The next three panels are the evolution of $\alpha$, $E_{\rm p}$, and energy Flux (1 - 10,000 keV), where orange represents the Band model results, and blue represents the CPL model results.
    {The slashed region in each panel indicates the time interval of TTE data losses.}
    }
    \label{fig:LC_para}
\end{figure}

We performed further analysis on the GBM data by using {\tt HEtools} (see in APPENDIX \ref{app:1}).
Before this tool was named, we have already applied it in some GRB data analysis \cite[e.g.,][]{jin2023detection,ren2022very,wang2022grb}.
The Fermi-GBM \citep{meegan2009fermi} payload has two types of detectors, including the 12 sodium iodide (NaI) detectors and 2 bismuth germanate (BGO) detectors.
The selection of detectors is usually based on the count rate and pointing direction, here we selected one NaI (na) detctor and one BGO (b1) detector for data analysis. 
{It is worth noting that the Fermi-GBM team reported that the Bad time interval due to the pulse pile-up is $T_0 + [3.00,~7.00]~\rm s$ ($T_0$ is the GBM trigger time) for the TTE data \citep{2023GCN.33551....1D}, as marked in Figure \ref{fig:LC_para}. The default time bin of CSPEC data is 1.024 s, while the rebin TTE data can yield spectrum with higher time resolution. And the time interval of pile-up of TTE data is shorter than that of CSPEC data. Therefore, in this work we take the TTE data for analysis, and extract data via \texttt{GBM Data Tools} \citep{GbmDataTools}}.
The red solid line in the first panel is the time interval given by the Bayesian block technique \citep{scargle2013studies} that can be used as a reference for time-resolved spectral time intervals. 

\subsection{Spectral Analysis}\label{sec:spec_ana}
Due to the extremely abundant photon count of this burst, we can perform very high-time-resolution spectrum analysis. 
{Excluding the Bad time interval}, a total of {88} intervals were selected for time-resolved spectral analysis, and the specific time intervals are shown in Table \ref{tab:1}.

We used two type of photon spectrum model in our spectral inference, the first is a smoothly joined broken power-law function (the so-called ``Band” function; \citealt{band1993batse}).
The Band function is written as
\begin{equation}
    N(E)=
    \begin{cases}
        A \big(\frac{E}{100\,{\rm keV}}\big)^{\alpha}{\rm exp}{\big(-\frac{E}{E_0}\big)}, \mbox{if $E<(\alpha-\beta)E_{0}$ }\\
        A\big[\frac{(\alpha-\beta)E_0}{100\,{\rm keV}}\big]^{(\alpha-\beta)}{\rm exp}{\big(\beta-\alpha\big)}\big(\frac{E}{100\,{\rm keV}}\big)^{\beta},
        \mbox{if $E > (\alpha-\beta)E_{0}$}
    \end{cases}
    \label{eq:band}
\end{equation}
where \emph{A} is the normalization constant, \emph{E} is the energy in unit of keV, $\alpha$ is the low-energy spectral index, $\beta$ is the high-energy photon spectral index, and \emph{E$_{0}$} is the {break} energy in the spectrum.
The peak energy in the $\nu F_\nu$ spectrum $E_{\rm p}$ is equal to $E_{0}\times(2+\alpha)$.
When the detection energy range is narrow or the high-energy photon count rate is low, the $\beta$ of the Band function is often not well constrained, so another model is a cutoff power-law function (CPL), written as
\begin{equation}
    { N(E)=A(\frac{E}{100\,{\rm keV}})^{\alpha}{\rm exp}(-\frac{E}{E_{\rm c}}) },
\end{equation}
where \emph{$\alpha$} is the power law photon spectral index, \emph{E$_{\rm c}$} is the {break} energy in the spectrum,
and the peak energy $E_{\rm p}$ is equal to $E_{\rm c}\times(2+\alpha)$.
For the possible components of the photosphere, we consider a multicolor blackbody (mBB) model \citep{ryde2010identification,hou2018multicolor} to describe it, which is
\begin{equation}
    N(E)=\frac{8.0525(m+1)K}{\Big[\big(\frac{T_{\rm max}}{T_{\rm min}}\big)^{m+1}-1\Big]}\Big(\frac{kT_{\rm min}}{\rm keV}\Big)^{-2}I(E),\label{N(E)}
\end{equation}
where
\begin{equation}
    I(E)=\Big(\frac{E}{kT_{\rm min}}\Big)^{m-1}\int_{\frac{E}{kT_{\rm max}}}^{\frac{E}{kT_{\rm min}}}\frac{x^{2-m}}{e^x-1}dx,\label{I(E)}
\end{equation}
where $x=E/kT$, the temperature range from $kT_{\rm min}$ to $kT_{\rm max}$, and the index $m$ of the temperature determines the shape of spectra.
In addition, the thermal component is usually accompanied by a non-thermal component, which is a power-law (PL) model with $\gamma$ index, written as $N(E) = A~E^{\hat{\gamma}}$.

Then we employ the Bayesian inference \citep{thrane2019introduction,van2021bayesian} for parameter estimation and model comparison.
We use {\tt Dynesty} \citep{speagle2020dynesty,skilling2006nested,higson2019dynamic} from the {\tt Bilby} \citep{ashton2019bilby} package as the posterior parameter sampler.
Usually for GBM data, the likelihood function used in Bayesian inference is {\tt pgstat}\footnote{\url{https://heasarc.gsfc.nasa.gov/xanadu/xspec/manual/XSappendixStatistics.html}}.
When considering different hypotheses with the same prior volume, model selection can be done by comparing Bayes factors. 
The Bayes factor (BF) is the ratio of the Bayesian evidence ($\mathcal{Z} = \int \mathcal{L}(d|\theta) \pi(\theta) d\theta$) for different models. 
The log of Bayes factor can be written as
\begin{equation}
    \ln\text{BF}^\text{A}_\text{B} = \ln({\cal Z}_\text{A}) - \ln({\cal Z}_\text{B}).
    \label{eq:7}
\end{equation}
When $\ln{\rm BF} > 8$, we can say that there is a ``strong evidence" in favor of one hypothesis over the other \citep{thrane2019introduction}.	

The posterior parameters and model selection of each model are shown in Table \ref{tab:1}.
As shown in Figure \ref{fig:LC_para}, the three panels at the bottom are the evolution of model parameters and energy flux ($1 - 10,000$ keV) over time, in which the blue points are the CPL model parameters, and the orange points are the Band model parameters.
It is worth noting that in the early phase (before $\sim$ 3 s) of this burst, the low-energy spectral index obtained by both the Band model and the CPL model exceed the synchrotron limit, also known as the ``Line of Death" \citep{preece1998synchrotron,preece2002consistency}.
Beside, all high-energy photon spectral indexes also exceed typical values ($\beta \sim-2$) \citep{preece2000batse},
which most likely corresponds to the exponential decay of the Planck function at the highest temperature or insufficient high-energy photon count rate.

\begin{figure}[t]
    \centering
    \includegraphics[width=0.32\textwidth]{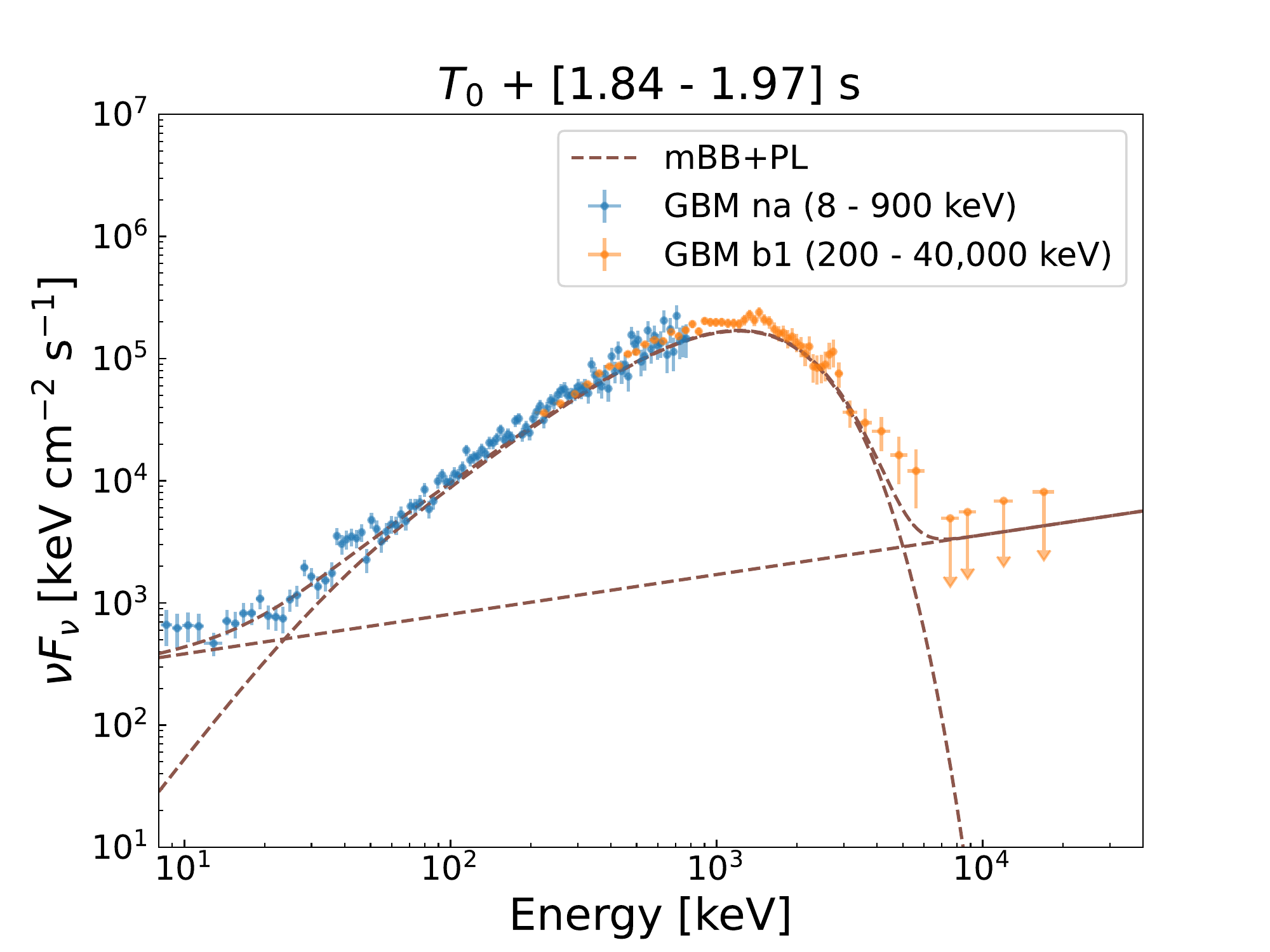}
    \includegraphics[width=0.32\textwidth]{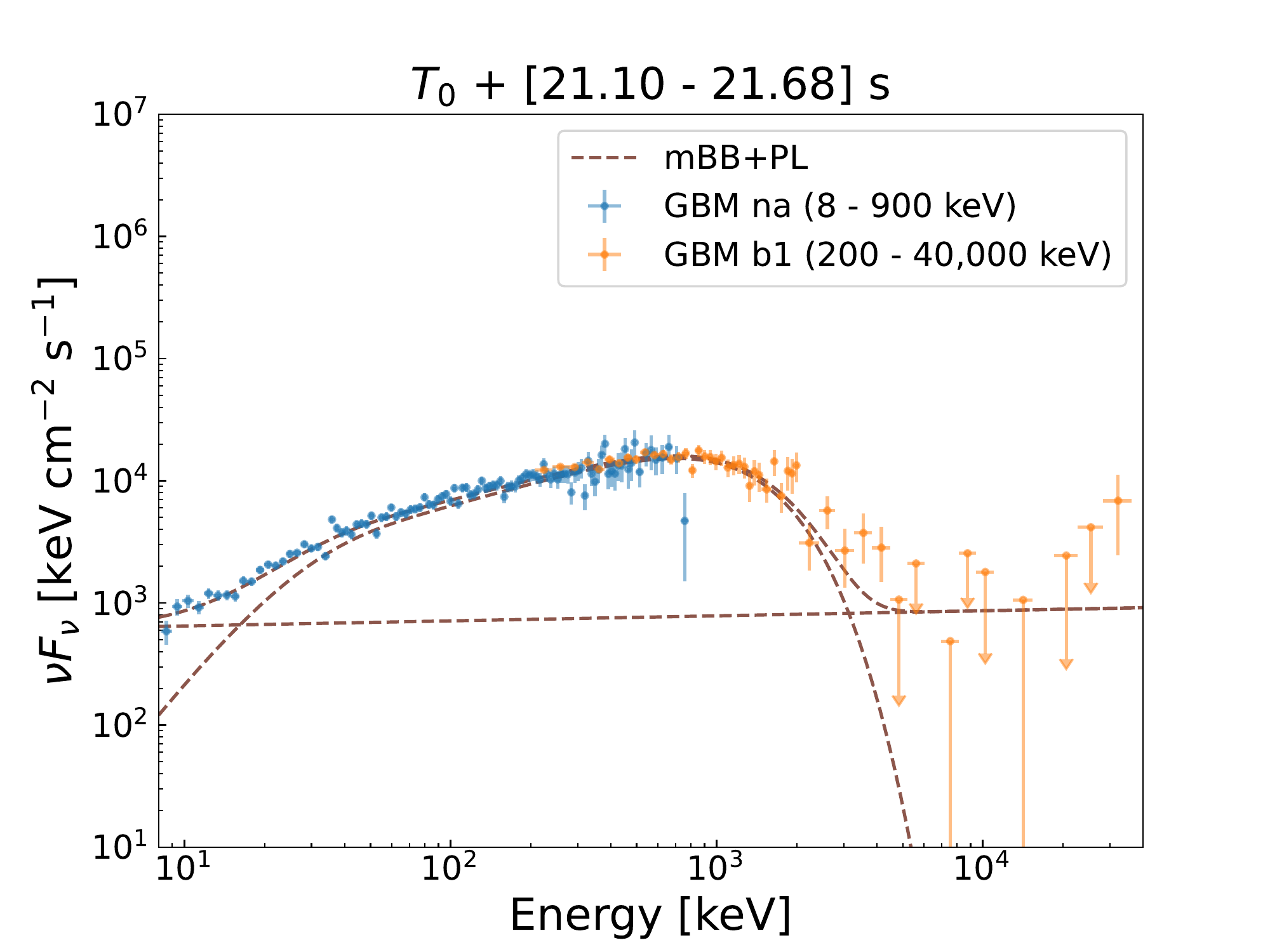}
    \includegraphics[width=0.32\textwidth]{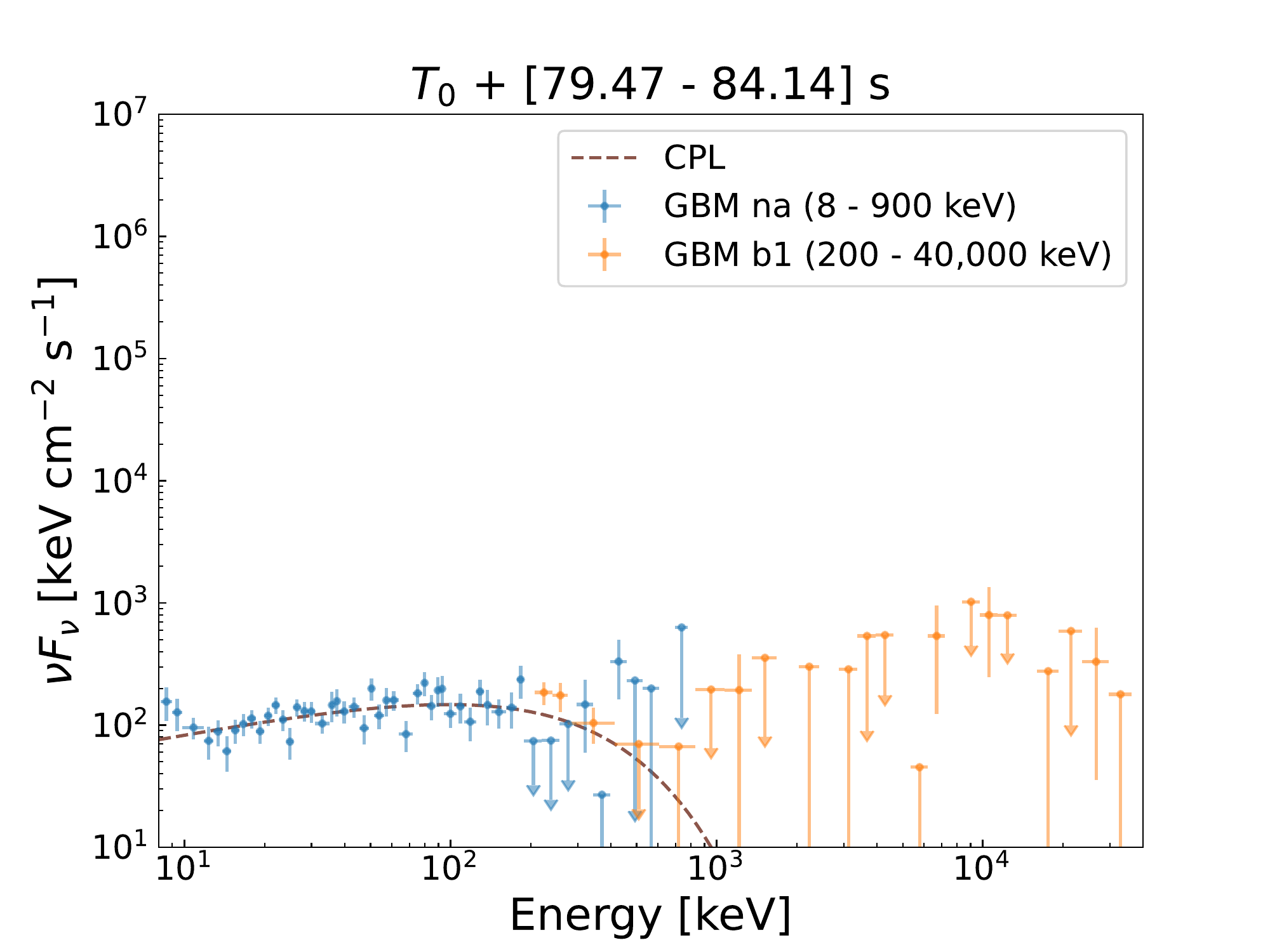}
    \includegraphics[width=0.32\textwidth]{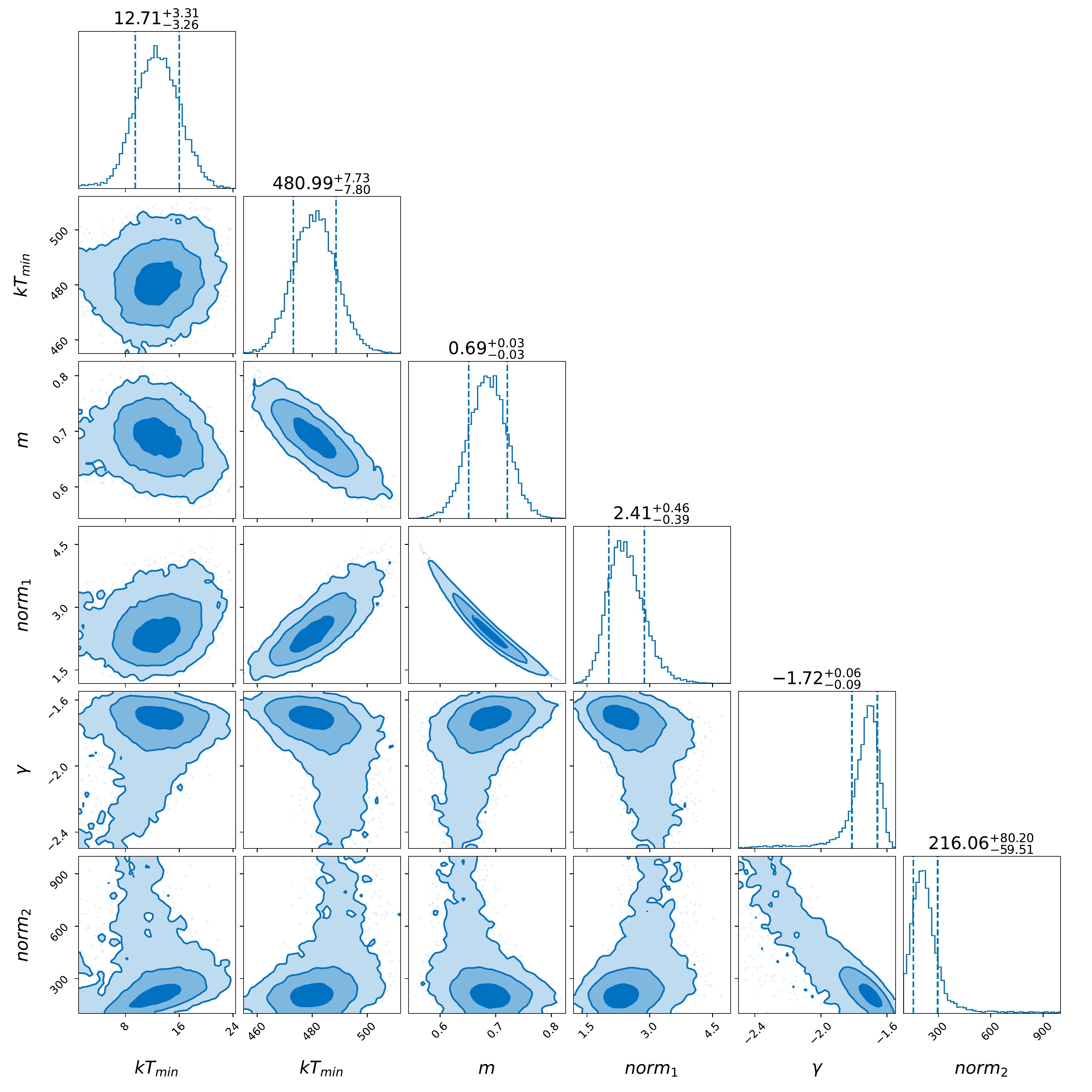}
    \includegraphics[width=0.32\textwidth]{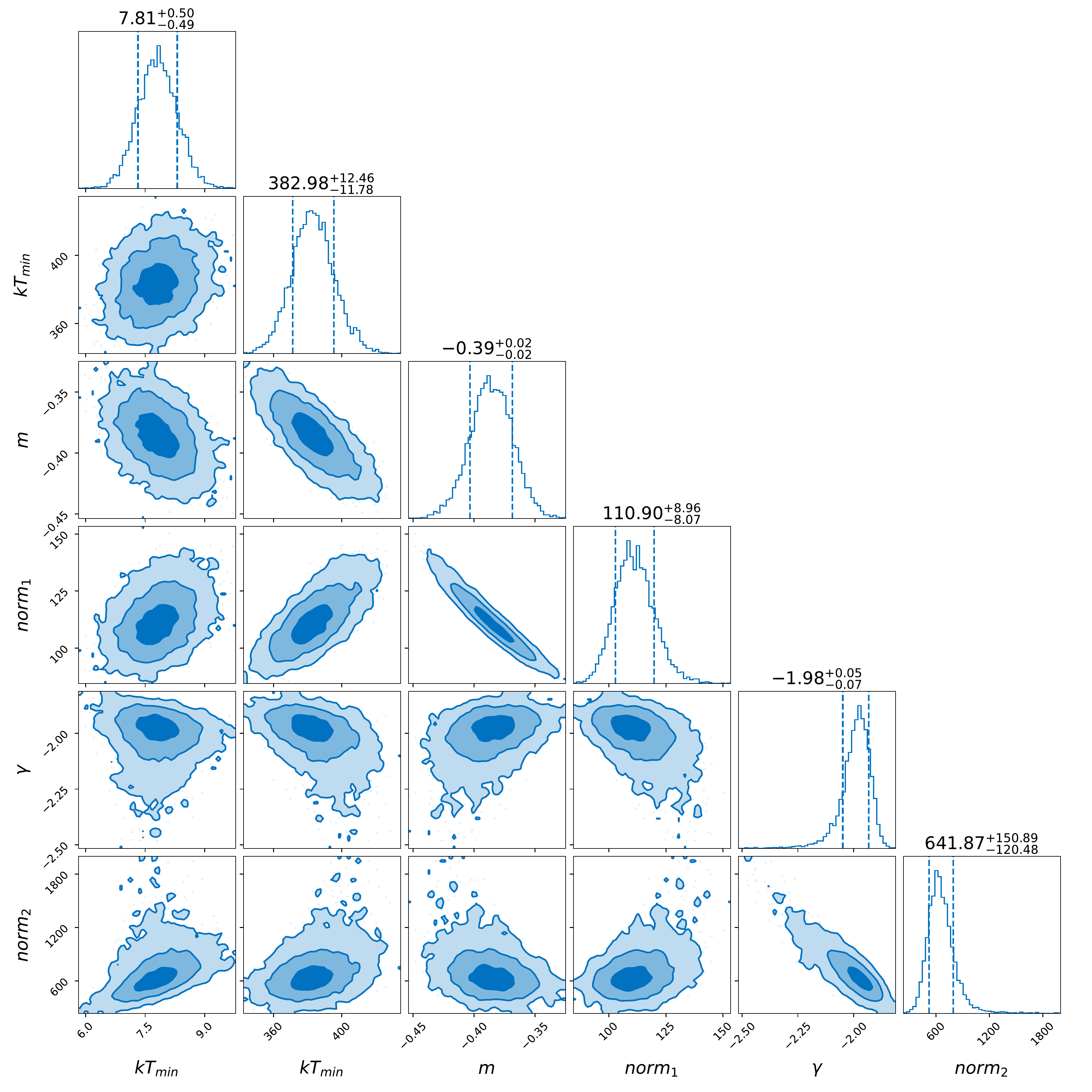}
    \includegraphics[width=0.32\textwidth]{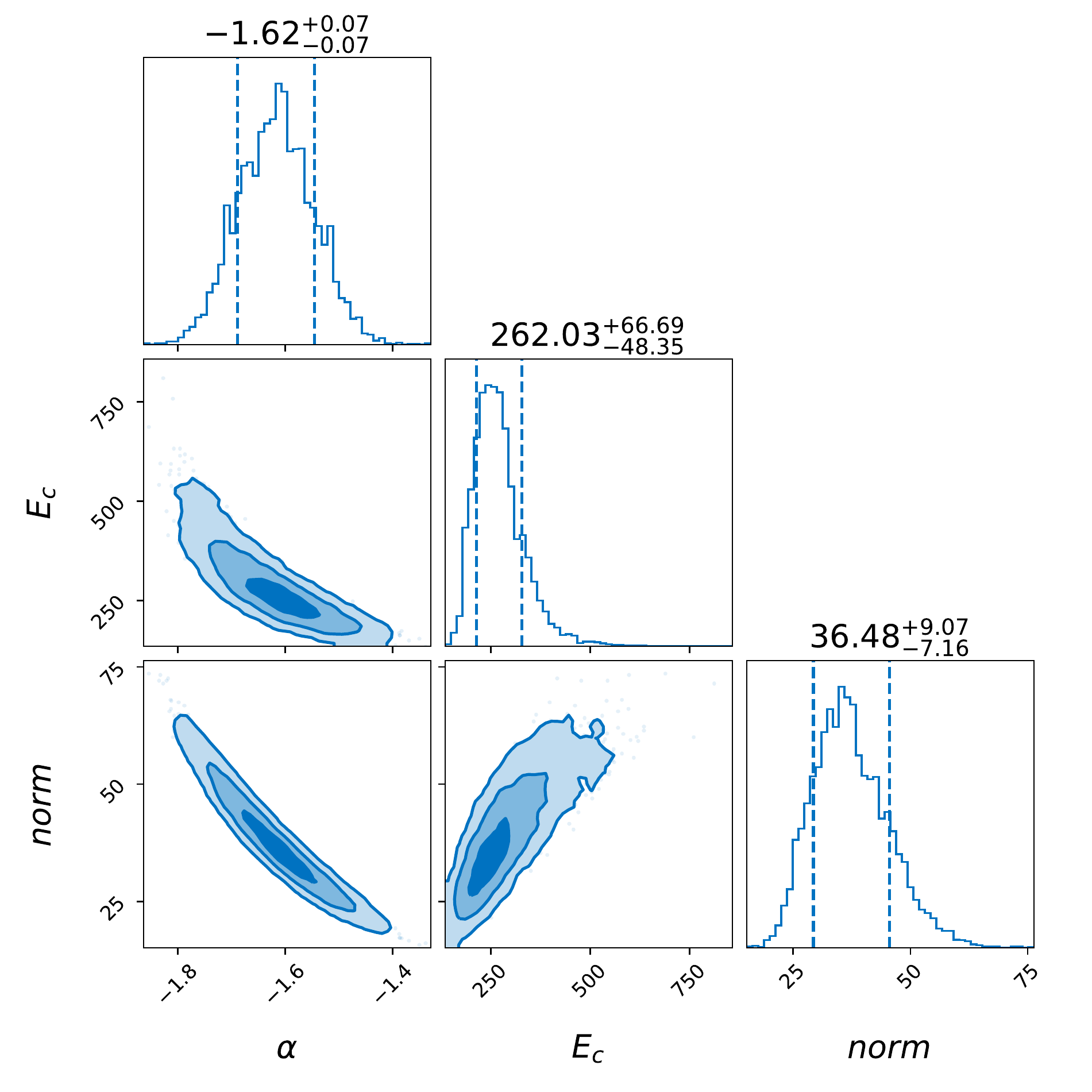}
    \caption{The $\nu F_{\nu}$ spectra of GRB~230307A at three special time intervals. From left to right in the first panel, the photon spectrum used in the first two time intervals is the mBB+PL model, and the last one is the CPL model.
    {The distributions of posterior parameters for each model are presented in the bottom panels.}}
    \label{fig:th_veo}
\end{figure}

The evolution of $E_{\rm p}$ and $\alpha$ over the entire outburst simultaneously shows the pattern of intensity tracking \citep{lu2012comprehensive,Ryde2019intensity}, also known as ``Double-tracking" \citep{li2019double}.
Even more peculiarly, we found a broken behavior in the $\alpha-F$ relation of GRB~230307A and called it broken ``$\alpha$-intensity".
As shown in Figure \ref{fig:th_veo}, we performed additional spectral analyzes for the $\alpha$-hardest interval ($T_0$ + [1.84 - 1.97] s), the broken-$\alpha$ interval ($T_0$ + [21.10 - 21.68] s), and a late interval ($T_0$ + [79.47 - 84.14] s). 
{These three time slices are outside the Bad time interval of the GBM TTE data.}
The $E_{\rm p}$ of the first time interval may come from the maximum temperature ($kT_{\rm max} \sim 480$ keV) of the mBB spectrum, and is accompanied by a non-thermal PL component ($\hat{\gamma} \sim -1.72$).
In the second time interval, the ratio of these two components (thermal vs. non-thermal) changed significantly compared to before. By the final time interval, the superposition of these two components has become indistinguishable, and the slope of the energy spectrum ($\alpha \sim -1.59$) is represented by the non-thermal emission.

\section{Characteristics}\label{sec:2}
\subsection{The $\alpha-F$ and $F-E_{\rm p}$ relations}
\begin{figure}[!htb]
    \centering
    \includegraphics[width=0.49\textwidth]{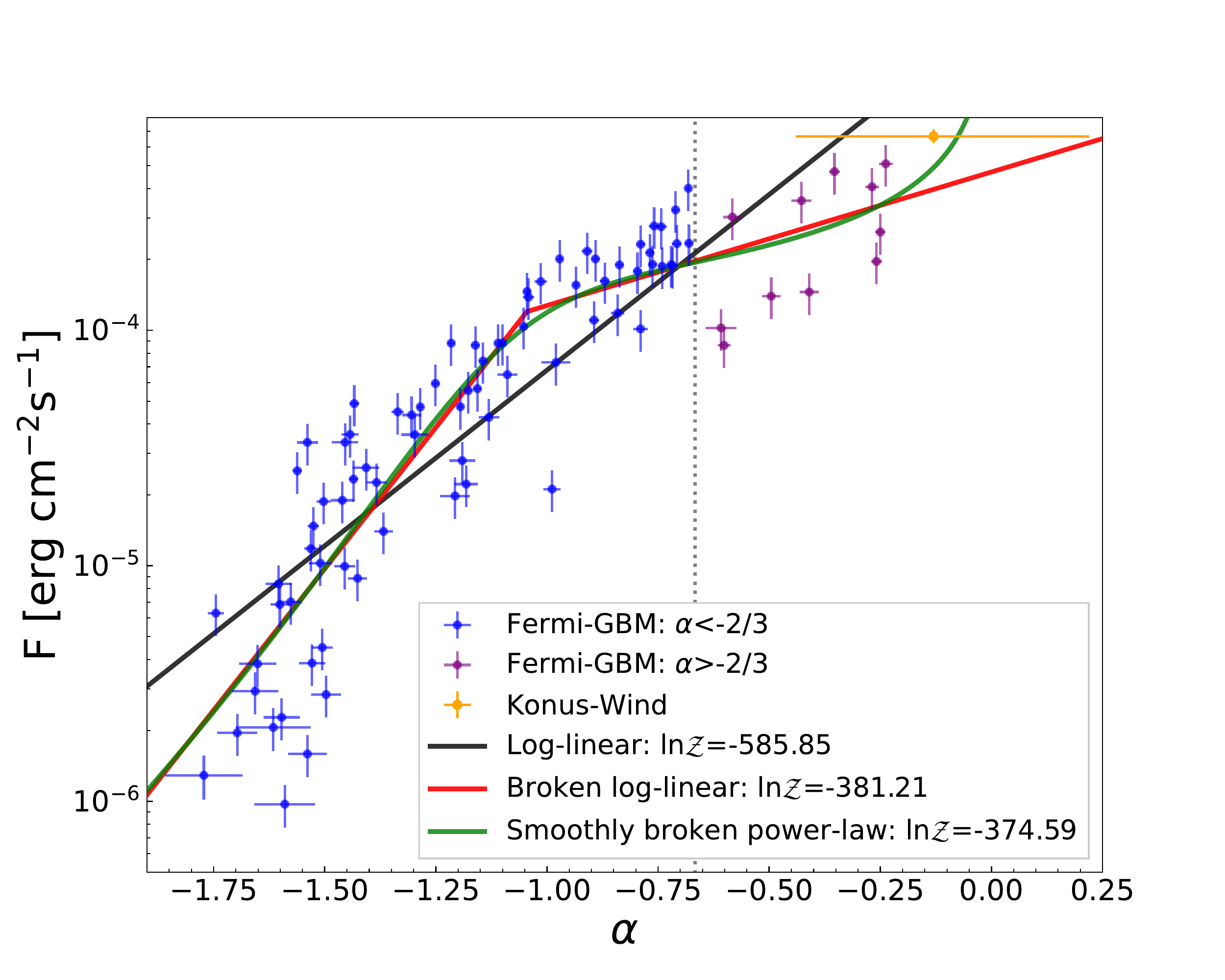}
    \includegraphics[width=0.49\textwidth]{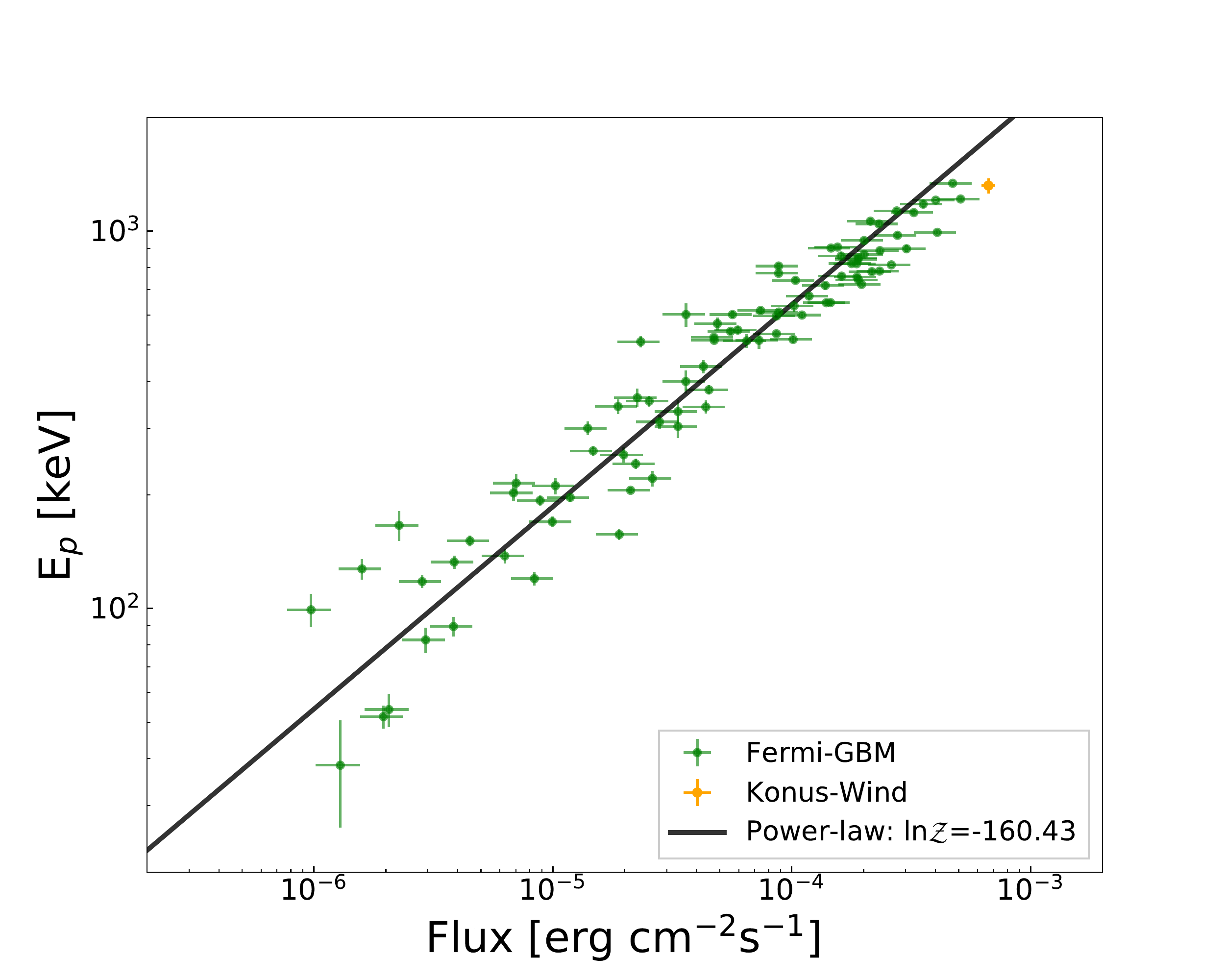}
    \caption{The $\alpha-F$ relation (left panel) and the $F-E_{\rm p}$ relation (right panel) for the CPL-model samples of GRB~230307A.
    In the left panel: The Fermi-GBM samples with the $\alpha$ exceeding the synchrotron limit ($\alpha=-2/3$, the grey dashed line) are marked in purple, and the others are in blue. The Konus-Wind data point, measured at the peak time, is marked in orange. The black/red/green line corresponds to the best-fit Log-linear/Broken Log-linear/Smoothly broken power-law model for the $\alpha-F$ relation, respectively.
    In the right panel: The black line corresponds to the best-fit Power-law model we obtained for the $\alpha-F$ relation.}
    \label{fig:afreation}
\end{figure}

Based on results obtained in Section \ref{sec:spec_ana}, we performed a statistical analysis on the two relations (i.e., $\alpha-F$, $F-E_{\rm p}$) of GRB~230307A for the CPL-model and Band-model samples (listed in the Table~\ref{tab:1}), respectively.
In order to be conservative, we consider 20\% uncertainty of the GBM effective area as the systematic error of energy flux $F$~\citep{2009ApJ...702..791M}.
{Since the GBM TTE data in the Bad time interval were excluded, we took the spectral properties of the peak emission measured by Konus-Wind, including the spectral index of $-0.13_{-0.31}^{+0.35}$, the flux of $6.66_{-0.42}^{+0.46}\times 10^{-4}~\rm erg/cm^2/s$, and the peak energy of $1321_{-62}^{+60}~\rm keV$ \citep{2023GCN.33427....1S}.  All of these data points are presented in Figure \ref{fig:afreation}.}

As for the $\alpha-F$ relation,
\cite{Ryde2019intensity} has analyzed the sample in \cite{yu2019Bayesian} and organized this relation as the log-linear (LL) function:
\begin{equation}
\label{eq:ll}
    F(\alpha) =  N~e^{k\alpha},
\end{equation}
where $N$ is normalization factor, and they found the parameter $k$ was about 3.
{Here we fit the log-linear relation function for the GRB~230307A and obtain the best-fit parameter $k$ of 3.43/3.43 for the CPL/Band-model samples with the corresponding log of Bayesian evidence $\ln({\cal Z}) = $ -585.85/-594.52, respectively.}
However, the $\alpha-F$ samples of GRB~230307A exhibit an obvious broken behavior as shown in the left panel of the Figure \ref{fig:afreation}.
In our work, we take two other relation function to fit the $\alpha-F$ samples:
One is the broken log-linear (BLL) function given as
\begin{equation}
\label{eq:bll}
    F(\alpha) = 
    \begin{cases}
    N~{\rm e}^{k_1 \alpha}, ~~\mbox{if $\alpha~<~\alpha_b$ }\\
    N'~{\rm e}^{k_2 \alpha}, ~~\mbox{if $\alpha~>~\alpha_b$ }\\
    \end{cases}
\end{equation}
where $N' = N~{\rm e}^{(k_1-k_2)\alpha_b}$, 
and {we get that the best-fit broken point $\alpha_b$ is $-1.05/-1.03$, the best-fit first index $k_1$ is $5.53/5.51$, and the best-fit second index $k_2$ is 1.31/1.23, corresponding to $\ln({\cal Z}) =-381.21/-383.62$ for the CPL/Band-model samples, respectively.}
The other is the smoothly broken power-law (SBPL) function written as
\begin{equation}
\label{eq:sbpl}
    F(\alpha) =  N~\bigg[\bigg(\frac{\alpha}{\alpha_b}\bigg)^{\gamma_1} + \bigg(\frac{\alpha}{\alpha_b}\bigg)^{\gamma_2} \bigg]^{-1}.
\end{equation}
{Based on the CPL/Band-model samples, the best-fit broken point $\alpha_b$ we obtained is $-1.14/-1.14$, the best-fit first index $\gamma_1$ is $9.52/9.60$, and the best-fit second index $\gamma_2$ is $0.56/0.57$, with the evidence of $\ln({\cal Z})= $  $-374.59/-380.74$.
Compared with the previous log-linear relation function, the broken log-linear and smoothly broken power-law functions exhibit large Bayes factors with $\ln{\rm BF}_{\rm LL}^{\rm BLL} = 204.64$ and $\ln{\rm BF}_{\rm LL}^{\rm SBPL} = 211.26$ for the CPL-model samples ($\ln{\rm BF}_{\rm LL}^{\rm BLL} = 210.90$ and $\ln{\rm BF}_{\rm LL}^{\rm SBPL} = 213.78$ for the Band-model samples), which indicates that there is a strong evidence of the broken ``$\alpha$-intensity" relation in GRB~230307A.}
The results of the $\alpha-F$ relation for the CPL-model sample are displayed in the left panel of the Figure \ref{fig:afreation}.

The $F-E_{\rm p}$ relation is found in a large fraction of GRBs and also in the time-resolved spectrum of single GRB \citep{wei2003there,yonetoku2004gamma,liang2004luminosity,lu2012comprehensive}, which can be naturally explained by the photosphere model \citep{fan2012photospheric}.
As for GRB~230307A , the $F-E_{\rm p}$ relation exhibits a linear relation in logarithmic space,  as shown in the right panel of Figure \ref{fig:afreation}.
This behavior coincides with the intensity trace in the third panel of Figure \ref{fig:LC_para}, which can be described as a single power-law function:
\begin{equation}
    E_{\rm p} = N~F^{\gamma}.
\end{equation}
where $N$ is the normalization factor and $\gamma$ is the flux index. 
{The best-fit index for GRB~230307A is $\gamma = 0.54/0.54$ based on the CPL/Band-model samples, respectively.} 
Our result on the $F-E_{\rm p}$ relation for the CPL-model samples is displayed in the right panel of the Figure \ref{fig:afreation}.

\subsection{Spectral lag}\label{sec:spec_lag}
\begin{figure}[t]
    \centering
    \includegraphics[width=0.32\textwidth]{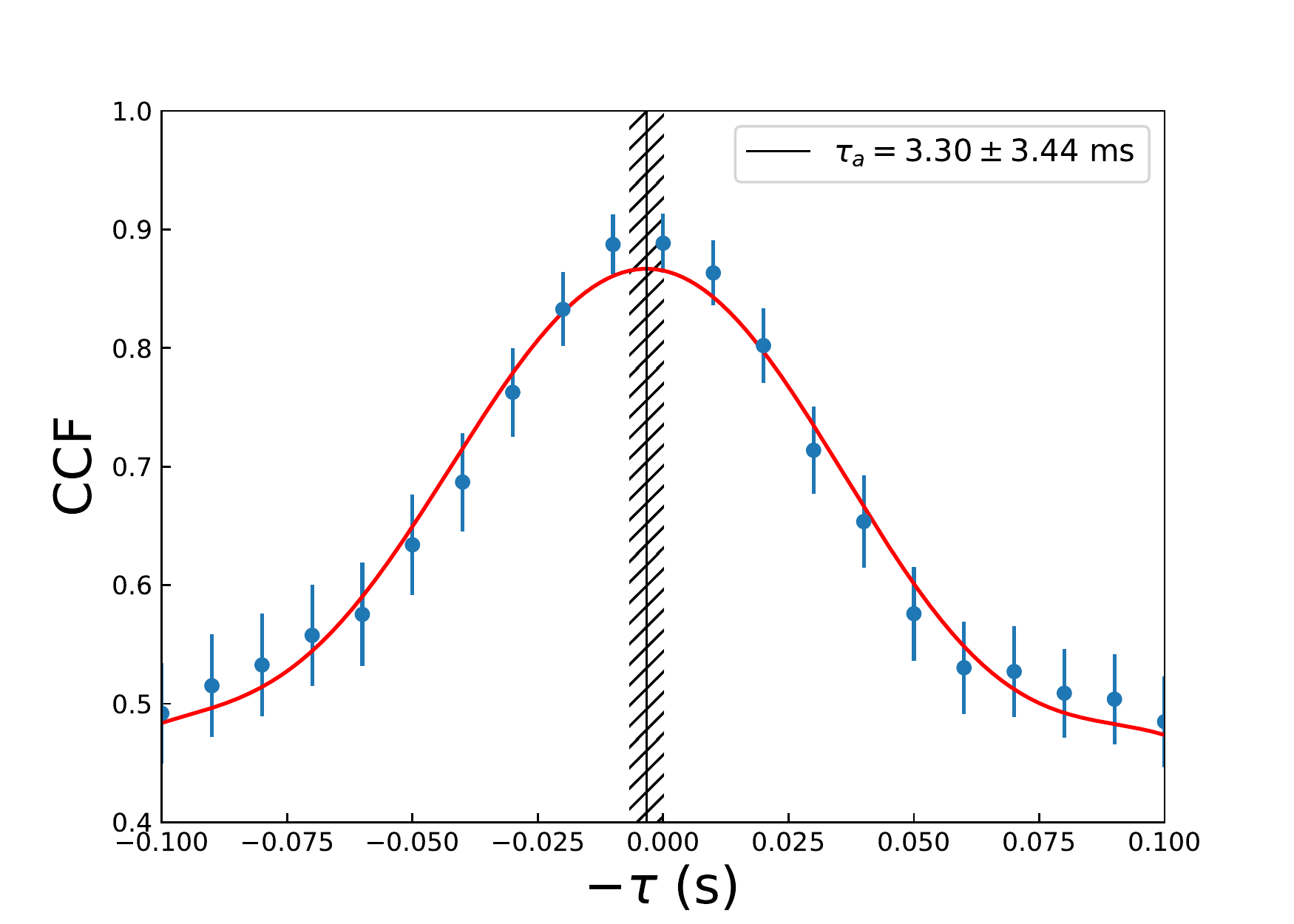}
    \includegraphics[width=0.32\textwidth]{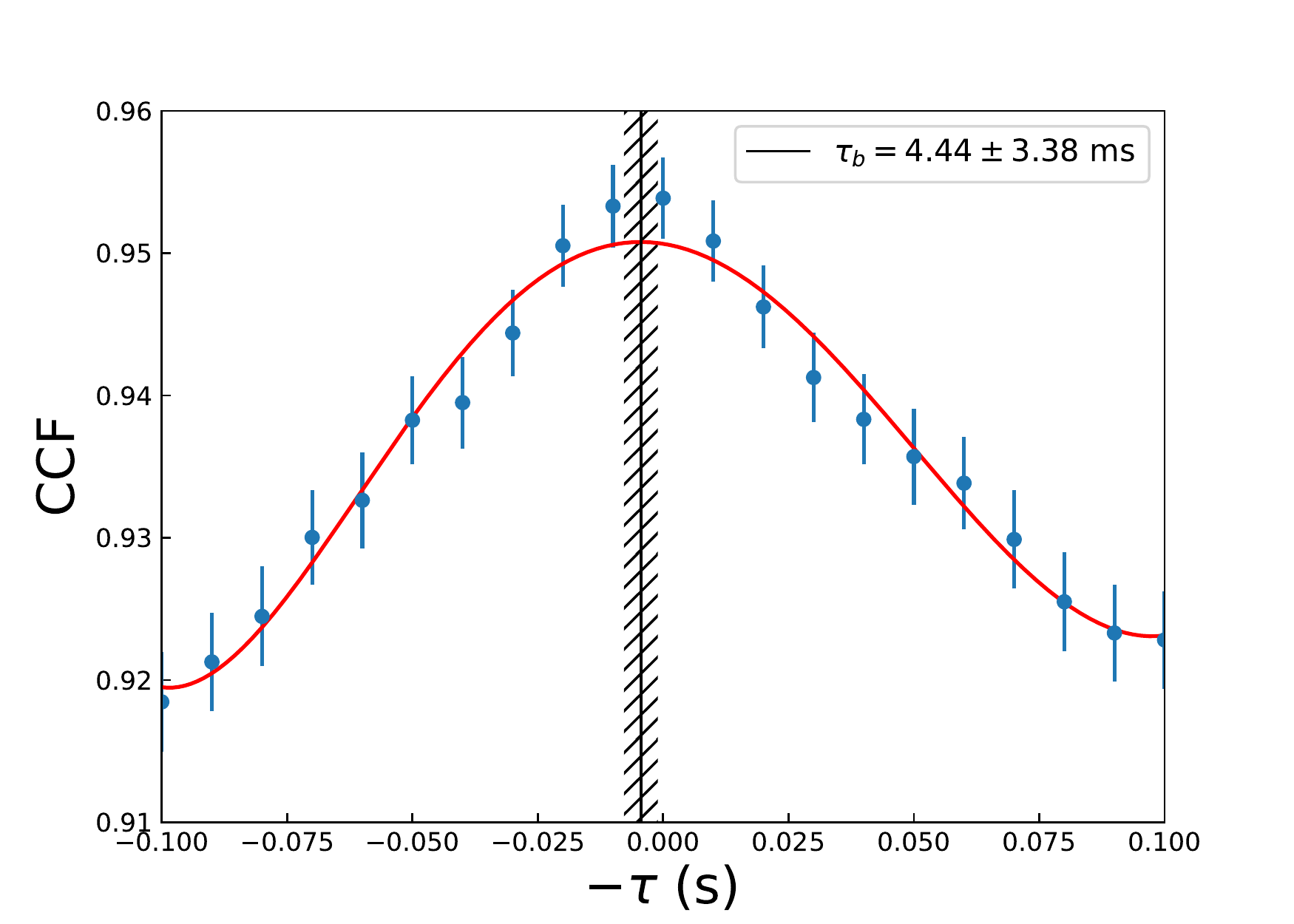}
    \includegraphics[width=0.32\textwidth]{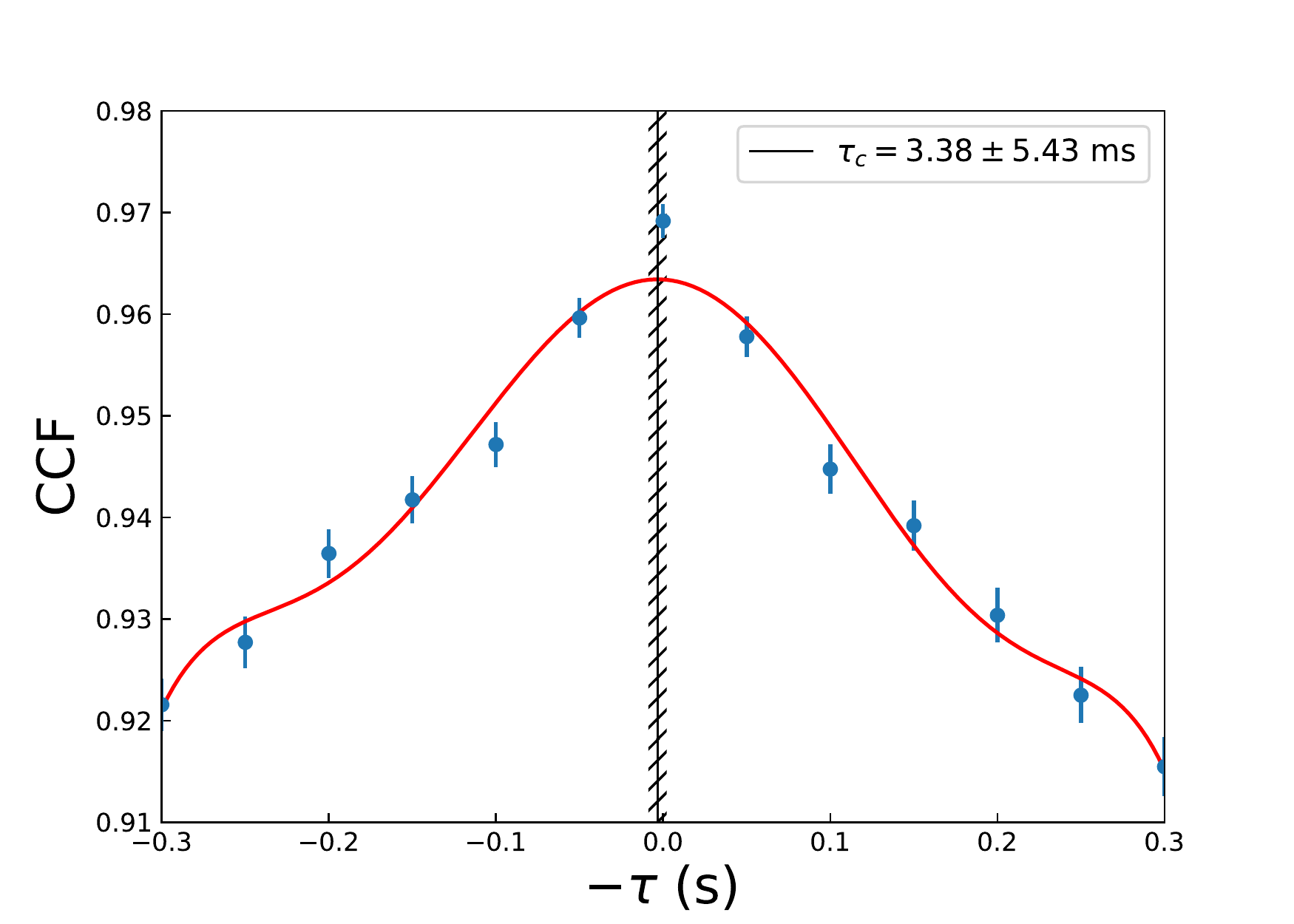}
    \caption{Spectral lags of three episodes. The time bin size of episode a and episode b is set to 0.01 s. And because the duration of episode b is longer, its time bin size is set to 0.05 s. The error comes from the Monte Carlo simulation of the light curve.}
    \label{fig:tau}
\end{figure}

Another important characteristic of GRBs is the spectral lag, in the energy range below 10 MeV, the high-energy photons arrive earlier than the low-energy photons \citep{norris2000connection,norris2002implications,norris2005long}.
The delay of pulse peaks in different energy bands can be quantified using the cross-correlation function (CCF), which is widely used in the calculation of GRB spectral lag \citep{band1997gamma,ukwatta2010spectral}.
{We calculated the CCF of the GRB~230307A time series in the energy band (100 - 150 keV) and (200 - 250 keV) with three different time intervals, episode a ($T_0+[0.2,~0.4]~\rm s$), episode b ($T_0+[0.4,~3.0]~\rm s$), episode c ($T_0+[7.0,~40.0]~\rm s$).
None of these episodes were affected by the GBM instrument pile-up. 
And we estimated the uncertainty of the lag by Monte Carlo simulation \citep{ukwatta2010spectral}.
The corresponding spectral lags are $\tau_a =  3.30 \pm 3.44 ~\rm{ms}$, $\tau_b = 4.44 \pm 3.38 ~\rm{ms}$, and $\tau_c = 3.38 \pm 5.43 ~\rm{ms}$, as shown in Figure \ref{fig:tau}. All are consistent with being zero. 
} 

\subsection{$E_{\rm p,z}$ - $E_{\gamma,\rm iso}$ relation}\label{sec:amati}
\begin{figure}[!h]
    \centering
    \includegraphics[width=0.75\textwidth]{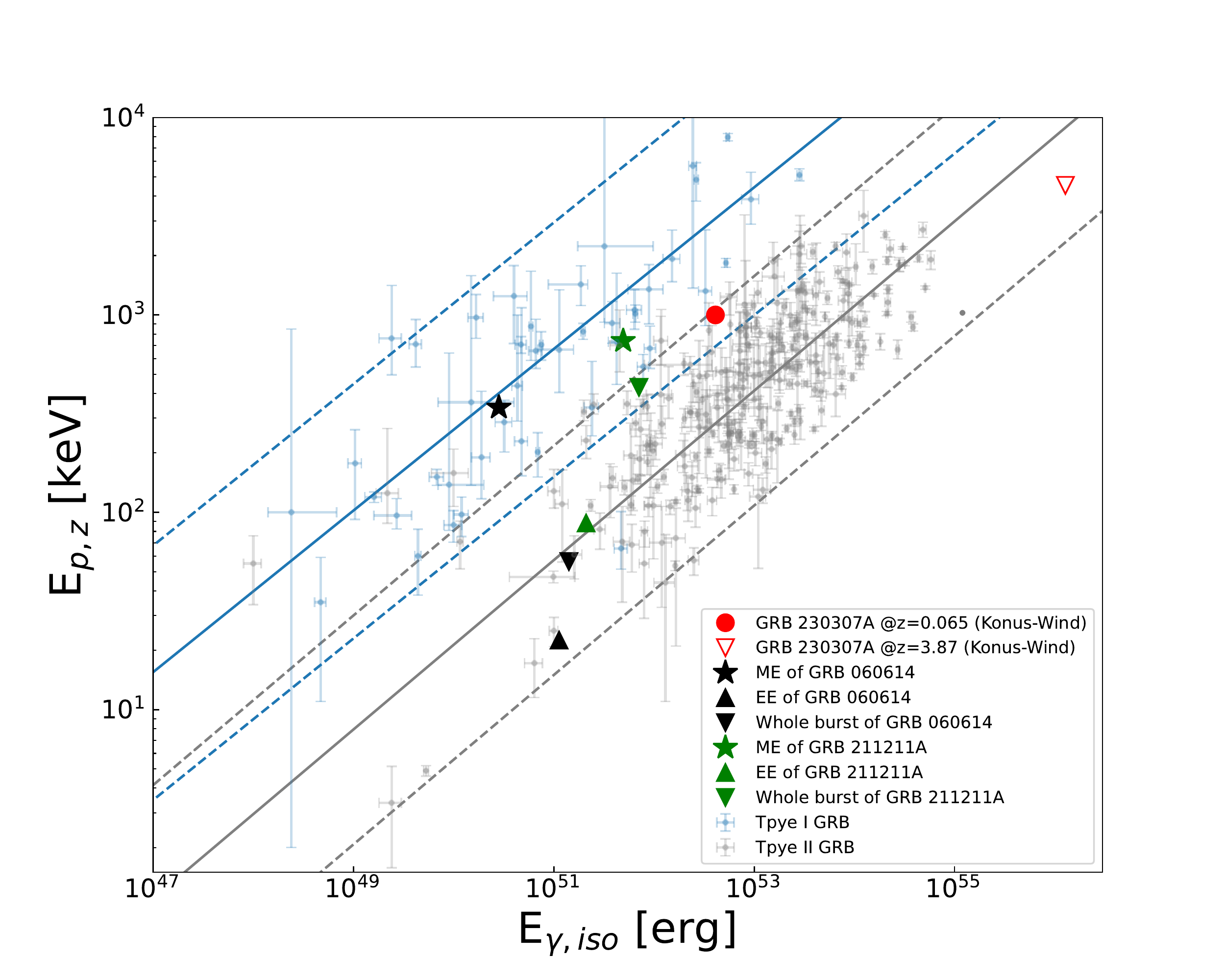}
    \caption{The $E_{\rm p,z}$ - $E_{\gamma,\rm iso}$ diagram.
    The blue and gray points are the data of Type I and Type II gamma-ray bursts with known redshifts, and the corresponding dashed lines represent the 2$\sigma_{\rm cor}$ correlation regions, respectively \citep{minaev2020p,minaev2020grb}. 
    Green and gold markers represent the different phases of GRB 060614, the first long duration burst (lasting about 100 seconds) with identified kilonova emission \citep{Yang2015,Jin2015,yang2022long}.
    The red circle and hollow inverted triangle are for GRB 230307A in the case of $z=0.065$ and $3.87$, respectively.
    }
    \label{fig:amati}
\end{figure}

The relation of $E_{\rm p,z}-E_{\gamma,\rm iso}$ \citep{amati2002intrinsic} is often used in the judgment of GRBs classification \citep[e.g.,][]{gehrels2006new}, where $E_{\rm p,z} = (1+z)E_{\rm p}$ is the rest frame peak energy, $E_{\gamma,\rm iso}$ is the isotropic bolometric emission energy, written as 
\begin{equation}
    E_{\gamma,\text{iso}} = \frac{4 \pi d_L^2 k S_{\gamma}}{1+z},
    \label{eq:E_gamma_iso}
\end{equation}
where $d_L$ is the luminosity distance, $S_{\gamma}$ is the energy fluence in the gamma-ray band, and $k$ is the correction factor, which can correct the energy range of the observer frame to the energy range of 1 - 10,000 keV in the rest frame. The correction factor $k$ \citep{bloom2001prompt} writes as
\begin{equation}
    k = \frac{\int_{1/(1+z)}^{10^4/(1+z)}  E N(E) {\rm d} E }{\int_{e_1}^{e_2} EN(E) {\rm d} E},
    \label{eq:k_cor}
\end{equation}
where $e_1$ and $e_2$ correspond to the energy range of the detector.
The redshift of GRB~230307A is assumed to be $z \sim 0.065$ \citep{GCN33485}, and the adopted cosmological parameters are \emph{H$_{0}$} = $\rm 69.6 ~k ms^{-1}~Mpc^{-1}$, $\Omega_{\rm m}= 0.29$, and $\Omega_{\rm \Lambda}= 0.71$. 
And the data of Type I and Type II GRBs with redshifts are adopted from \cite{minaev2020p}. Unlike GRB 060614 and GRB 211211A that well resemble the short GRBs (i.e., Type I), with the sole prompt emission properties, it is challenging to uniquely classify GRB 230307A (see Figure \ref{fig:amati}). 
In addition, we also consider the fluence and peak energy of the time-integrated spectrum reported by Konus-Wind \citep{2023GCN.33427....1S,svinkin2023grb}.
If GRB 230307A had a redshift of 3.87, as speculated in \citet{levan2023grb}, it would be a Type II event (see  the inverted triangle in  Figure \ref{fig:amati}) with a huge $E_{\gamma,\rm iso}\sim 10^{56}$ erg. 

\section{Discussion}\label{sec:3}
\subsection{The broken ``$\alpha$-intensity" relation}
For the GRB prompt emission, the low-energy spectral index manifested as intensity-tracking evolution pattern occur in a small number of samples.
\cite{Ryde2019intensity}  has organized the $\alpha-F$ relation as the log-linear ($F \propto e^{k\alpha}$) function, which is believed to be a manifestation of subphotospheric heating in a flow with a varying entropy \citep{Ryde2017Emission,Ryde2019intensity}.
However, $\alpha -$intensity was also expected to be picked from non-thermal, i.e., synchrotron radiation, dominated GRBs, such as GRB~131231A \citep{li2019double} and GRB~211211A \citep{yang2022long}.
The evolution of $\alpha$ in the synchrotron radiation scenario may originate from the change of the magnetic field at the emission region \citep[e.g.,][]{uhm2014fast}. 
There are no clear observations and studies pointing out the impact on the $\alpha-F$ relationship when the proportion of thermal and non-thermal components evolve.

Thanks to the sufficient photon count GRB~230307A, we discovered an interesting phenomenon through the high-time-resolution spectrum analysis.
In Figure \ref{fig:afreation}, by counting the parameters of 106 time-resolved spectra, we found that there is an obvious broken behavior exhibited in the $\alpha-F$ relation.
In the left panel of Figure \ref{fig:th_veo}, the energy spectrum of the $\alpha$-hardest time interval can be fitted by the mBB+PL model,
and there is a correlation between $F-E_{\rm p}$.
Although the synchrotron radiation model can predict the $F-E_{\rm p}$ relation, this relation is more natural in the photosphere model \citep{fan2012photospheric}, and we consider that the thermal component is dominant in the early phase of GRB~230307A.

When the proportion of thermal components gradually decreases (as shown in the middle panel in Figure \ref{fig:th_veo}), the $\alpha-F$ relation begins to show an broken behavior.
The weaker emission and the broader spectrum may come from the entropy decreases the photosphere secedes from the saturation radius \citep{Ryde2019intensity}. 
And the broken point $\alpha_b \sim -1.14 $ is consistent with the typical low-energy spectral index of GRBs \citep[$\alpha \sim -1$;][]{kaneko2006complete,preece2000batse}, 
which may arise from different physical situations \citep[e.g., transfer simulations of magnetized jets with $\epsilon_B > 10^{-3}$;][]{Vurm2016Radiative}, and can be summarized as unsaturated comptonization of low-energy photons.
One of the time-resolved spectra of this phase is shown in the middle panel of Figure \ref{fig:th_veo}, which may explain the statistically significant hidden black body components found on the left side of the Band model energy peak in some previous studies \citep{guiriec2011detection,axelsson2012grb110721a,guiriec2013evidence}.

While at a later phase, the spectral index of the CPL model $\alpha \sim -1.59$, and the emission may be dominated by non-thermal components (shown in the right panel of Figure \ref{fig:th_veo}).
Hence, GRB~230307A is a peculiar case which is recorded the detailed transition of the emission component from thermal to non-thermal, and characterizes a special broken ``$\alpha$-intensity" behavior.

\subsection{Comparison with other GRBs}
First we compare GRB~230307A with some long-short (also known as ``hybrid") bursts, i.e., the bursts with a duration much longer than 2 seconds but from the merger of the compact objects. 
Unlike GRB~060614, 
GRB~230307A can not be directly divided into two emission epochs (see Figure \ref{fig:LC_para}). In other words, it is not reasonable to attribute the long duration of GRB 230307A to a short hard spike followed by a soft long tail. Anyhow, for a redshift $z=0.065$, GRB~230307A is slightly closer to the region of Type I bursts than that of Type II in the $E_{\rm p,z}$ - $E_{\gamma,\rm iso}$ plane (see Figure \ref{fig:amati}). This burst also has a $\sim 0$ spectral lag in each episode, similar to GRB~060614,  GRB~211211A and other type I GRBs.
Beside, GRB~230307A is suggested to have a projected offset of $\sim$40 kpc from the candidate host galaxy \citep{GCN33485}, which is in favor of the neutron star binary progenitor. 
If this event is indeed outside its host galaxy, it would be easier to detect the  
Li-Paczynski macronova \citep[also known as the kilonova;][]{li1998transient} signal, which would be a ``smoking gun" signature for the compact-binary merger origin, as the case of GRB 070809 \citep{2020NatAs...4...77J}. Indeed, the kilonova signal likely has already been detected by the James Webb Space Telescope  \citep[JWST;][]{levan2023grb}. The successful detection of such a signal in the first ever visit of JWST to the possible merger event is strongly in favor of the long-standing speculation that the kilonovae are ubiquitous \citep{2016NatCo...712898J}. 
For the speculated  $z=3.87$, GRB~230307A would be a long GRB (i.e., type II) with a huge $E_{\gamma,\rm iso}\sim 10^{56}$ erg (as indicated by the red inverted triangle in Figure \ref{fig:amati}), which would require a peculiar origin \citep{levan2023grb}. Moreover, the JWST observations on 5 April 2023 would suggest a thermal-like component with a luminosity of $L_{\rm th}\sim  7\times 10^{43}~{\rm erg~s^{-1}}$ and an intrinsic temperature (i.e., measured in the rest frame of the host galaxy) of $T'\sim 4000$ K. {\it Similar optical outburst, characterized by a very high luminosity but a quite low temperature} at $t'=28.8~{\rm day}/(1+z)\approx 5.9$ day after the burst, {\it has never been detected before} \citep{2016NatAs...1E...2L} and would require a large radiation radius of $R_{\rm e}\sim 2\times 10^{16}~{\rm cm}~(L_{\rm th}/7\times 10^{43}~{\rm erg~s^{-1}})^{1/2}(T'/4000~{\rm K})^{-2}$. If the large radius is due to the expansion of a supernova-like outflow (i.e., the progenitor was a massive star), its velocity would be in the order of $c$, the speed of the light. The required outflow mass as well as the corresponding kinetic energy are too large to be realistic. Given the above facts, we think the high redshift possibility is less likely.

In the energy spectrum analysis, we found that there may be a photosphere emission component in the early phase of GRB~230307A, which is similar to that of the also very bright GRB~130427A \citep{preece2014first}. 
However, the Brightest Of All Time (the BOAT, GRB~221009A) is believed to be dominated by non-thermal synchrotron radiation throughout its duration \citep{yang2023synchrotron}.
So these bright GRBs clearly have different emission components.
A typical example of the evolution of emission components is GRB~160625B, which has three episodes dominated by different components \citep{zhang2018transition}. 
Since GRB~230307A is bright enough, and the transition from thermal to non-thermal emission is clear in the time-resolved spectra.

The brightest GRB~221009A has been reported to be a recurrence timescale about 10,000 years \citep{Burns2023BOAT}.
For a given GRB with observed fluence $S$, its annual rate can be estimated with $R_{\rm GRB} = 1.037\times 10^{-5} \times S^{-3/2}$,
and the corresponding recurrence timescale $\tau (S)= R_{\rm GRB}^{-1}$ \citep{Burns2023BOAT}.
GRB~20307A has a fluence of $S \sim 4.05\times 10^{-3} ~{\rm erg~cm^{-2}}$ \citep{svinkin2023grb}, and the derived $\tau (S) \sim 24.85$ yrs.
Although its event rate is much higher than GRB~221009A, as the possible second brightest GRB, it is also very rare in Fermi's entire orbital career.

\section{SUMMARY}\label{sec:4}
In this work, we use {\tt HEtools} to analyze the Fermi-GBM data of the possibly second brightest GRB~230307A, and its features can be summarized as follows:
\begin{itemize}
\item {GRB~230307A has a very high count rate and also a long duration for the high-time-resolution spectral analysis. 
Through the parameter statistics inferred from up to 88 resolved spectra, we found a clear broken behavior in the $\alpha-F$ relation with a corresponding highest log Bayes factor,
which may be interpreted as the transition from thermal emission to non-thermal emission, and the significant evolution of photosphere throughout its duration.}
    
\item {Different from GRB~060614 and GRB~211211A, the light curve of GRB 230307A  is not composed of a hard spike followed by a soft tail (see the upper panel of Fig.\ref{fig:LC_para}). Nevertheless, the spectral lag is consistent with being zero, and for a redshift $z=0.065$ GRB~230307A is slightly closer to the region of Type I bursts than that of Type II in the $E_{\rm p,z}$ - $E_{\gamma,\rm iso}$ plane (see Figure \ref{fig:amati}). Considering also the $\sim 40$ kpc offset from the host galaxy as well as the kilonova signal detected by JWST, {\it it is reasonable to attribute GRB~230307A to a neutron star merger}. If instead the speculated $z=3.87$ holds, GRB 230307A would have $E_{\rm \gamma,iso}\sim 10^{56}$ erg. Moreover, there should present a new kind of optical outbursts characterized by a very high luminosity ($\sim 10^{44}~{\rm erg~s^{-1}}$) but rather low intrinsic temperature ($T'\sim 4000$ K) at $t'=t/(1+z)\sim 6$ day after the burst. This is because though some super-luminous supernovae and tidal disruption events can yield such energetic optical/ultraviolet emission, their temperatures are much higher and their declines are much shallower \citep{2016NatAs...1E...2L}. {We thus conclude that the high redshift possibility is unlikely.}
}

\item{Although recurrence timescale {($\tau (S)=24.85$ yrs)} of this tiny monster is lower than that of the BOAT GRB~221009A, it remains very rare throughout the Fermi's lifetime.
}
    
\end{itemize}

\section*{Acknowledgments}
We appreciate the anonymous referee for the helpful suggestions and acknowledge the use of the Fermi archive's public data.
This work is supported by the Natural Science Foundation of China (NSFC) under grants of No. 11921003,  No. 11933010 and No. 12003069.

\software{\texttt{Matplotlib} \citep{Hunter}, \texttt{Numpy} \citep{harris2020array}, 
	\texttt{bilby} \citep{ashton2019bilby}, \texttt{GBM Data Tools} \citep{GbmDataTools}}

\bibliography{bibtex}

\begin{thebibliography}{}
\expandafter\ifx\csname natexlab\endcsname\relax\def\natexlab#1{#1}\fi

\bibitem[{Abdo {et~al.}(2009)Abdo, Ackermann, Ajello, Asano, Atwood, Axelsson,
  Baldini, Ballet, Barbiellini, Baring, {et~al.}}]{abdo2009fermi}
Abdo, A., Ackermann, M., Ajello, M., {et~al.} 2009, The Astrophysical Journal,
  706, L138

\bibitem[{Amati {et~al.}(2002)Amati, Frontera, Tavani, Antonelli, Costa,
  Feroci, Guidorzi, Heise, Masetti, Montanari, {et~al.}}]{amati2002intrinsic}
Amati, L., Frontera, F., Tavani, M., {et~al.} 2002, Astronomy \& Astrophysics,
  390, 81

\bibitem[{Ashton {et~al.}(2019)Ashton, H{\"u}bner, Lasky, Talbot, Ackley,
  Biscoveanu, Chu, Divakarla, Easter, Goncharov, {et~al.}}]{ashton2019bilby}
Ashton, G., H{\"u}bner, M., Lasky, P.~D., {et~al.} 2019, The Astrophysical
  Journal Supplement Series, 241, 27

\bibitem[{Axelsson {et~al.}(2012)Axelsson, Baldini, Barbiellini, Baring,
  Bellazzini, Bregeon, Brigida, Bruel, Buehler, Caliandro,
  {et~al.}}]{axelsson2012grb110721a}
Axelsson, M., Baldini, L., Barbiellini, G., {et~al.} 2012, The Astrophysical
  journal letters, 757, L31

\bibitem[{Band {et~al.}(1993)Band, Matteson, Ford, Schaefer, Palmer, Teegarden,
  Cline, Briggs, Paciesas, Pendleton, {et~al.}}]{band1993batse}
Band, D., Matteson, J., Ford, L., {et~al.} 1993, The Astrophysical Journal,
  413, 281

\bibitem[{Band(1997)}]{band1997gamma}
Band, D.~L. 1997, The Astrophysical Journal, 486, 928

\bibitem[{Bloom {et~al.}(2001)Bloom, Frail, \& Sari}]{bloom2001prompt}
Bloom, J.~S., Frail, D.~A., \& Sari, R. 2001, The Astronomical Journal, 121,
  2879

\bibitem[{{Burns} {et~al.}(2023{\natexlab{a}}){Burns}, {Goldstein}, {Lesage},
  {Dalessi}, \& Team}]{GCN33414}
{Burns}, E., {Goldstein}, A., {Lesage}, S., {Dalessi}, S., \& Team, F.~G.
  2023{\natexlab{a}}, GRB Coordinates Network, 33414, 1

\bibitem[{{Burns} {et~al.}(2023{\natexlab{b}}){Burns}, {Svinkin}, {Fenimore},
  {Ag{\"u}{\'\i} Fern{\'a}ndez}, {Frederiks}, {Kann}, {Hamburg}, {Lesage},
  {Temiraev}, {Tsvetkova}, {Bissaldi}, {Briggs}, {Fletcher}, {Goldstein},
  {Hui}, {Hristov}, {Kocevski}, {Lysenko}, {Mailyan}, {Racusin}, {Ridnaia},
  {Roberts}, {Ulanov}, {Veres}, {Wilson-Hodge}, \& {Wood}}]{Burns2023BOAT}
{Burns}, E., {Svinkin}, D., {Fenimore}, E., {et~al.} 2023{\natexlab{b}}, arXiv
  e-prints, arXiv:2302.14037

\bibitem[{Crider {et~al.}(1997)Crider, Liang, Smith, Preece, Briggs, Pendleton,
  Paciesas, Band, \& Matteson}]{crider1997evolution}
Crider, A., Liang, E., Smith, I., {et~al.} 1997, The Astrophysical Journal,
  479, L39

\bibitem[{Daigne {et~al.}(2011)Daigne, Bo{\v{s}}njak, \&
  Dubus}]{daigne2011reconciling}
Daigne, F., Bo{\v{s}}njak, {\v{Z}}., \& Dubus, G. 2011, Astronomy \&
  Astrophysics, 526, A110

\bibitem[{{Dalessi} \& {Fermi GBM Team}(2023)}]{2023GCN.33551....1D}
{Dalessi}, S., \& {Fermi GBM Team}. 2023, GRB Coordinates Network, 33551, 1

\bibitem[{{Dalessi} {et~al.}(2023){Dalessi}, {Roberts}, {Meegan}, \&
  Team}]{GCN33411}
{Dalessi}, S., {Roberts}, O.~J., {Meegan}, C., \& Team, F.~G. 2023, GRB
  Coordinates Network, 33411, 1

\bibitem[{Deng \& Zhang(2014)}]{deng2014low}
Deng, W., \& Zhang, B. 2014, The Astrophysical Journal, 785, 112

\bibitem[{Fan {et~al.}(2022)Fan, Chang, Guo, Yuan, Hu, Li, Yue, Huang, Liu,
  Feng, {et~al.}}]{fan2022very}
Fan, Y., Chang, J., Guo, J., {et~al.} 2022, Acta Astronomica Sinica, 63, 27

\bibitem[{Fan {et~al.}(2012)Fan, Wei, Zhang, \& Zhang}]{fan2012photospheric}
Fan, Y.-Z., Wei, D.-M., Zhang, F.-W., \& Zhang, B.-B. 2012, The Astrophysical
  Journal Letters, 755, L6

\bibitem[{{Fermi GBM Team}.(2023)}]{GCN33405}
{Fermi GBM Team}. 2023, GRB Coordinates Network, 33405, 1

\bibitem[{Gehrels {et~al.}(2006)Gehrels, Norris, Barthelmy, Granot, Kaneko,
  Kouveliotou, Markwardt, M{\'e}sz{\'a}ros, Nakar, Nousek,
  {et~al.}}]{gehrels2006new}
Gehrels, N., Norris, J., Barthelmy, S., {et~al.} 2006, Nature, 444, 1044

\bibitem[{Ghirlanda {et~al.}(2003)Ghirlanda, Celotti, \&
  Ghisellini}]{ghirlanda2003extremely}
Ghirlanda, G., Celotti, A., \& Ghisellini, G. 2003, Astronomy \& Astrophysics,
  406, 879

\bibitem[{{Gillanders} {et~al.}(2023){Gillanders}, {O'Connor}, {Dichiara},
  {Troja}, \& behalf of a~larger team:}]{GCN33485}
{Gillanders}, J., {O'Connor}, B., {Dichiara}, S., {Troja}, E., \& behalf of
  a~larger team:. 2023, GRB Coordinates Network, 33485, 1

\bibitem[{Goldstein {et~al.}(2021)Goldstein, Cleveland, \&
  Kocevski}]{GbmDataTools}
Goldstein, A., Cleveland, W.~H., \& Kocevski, D. 2021, Fermi GBM Data Tools:
  v1.1.0, ,

\bibitem[{Golenetskii {et~al.}(1983)Golenetskii, Mazets, Aptekar, \&
  Ilyinskii}]{golenetskii1983correlation}
Golenetskii, S., Mazets, E., Aptekar, R., \& Ilyinskii, V. 1983, Nature, 306,
  451

\bibitem[{Goodman(1986)}]{goodman1986gamma}
Goodman, J. 1986, The Astrophysical Journal, 308, L47

\bibitem[{Guetta {et~al.}(2001)Guetta, Spada, \& Waxman}]{guetta2001efficiency}
Guetta, D., Spada, M., \& Waxman, E. 2001, The Astrophysical Journal, 557, 399

\bibitem[{Guiriec {et~al.}(2011)Guiriec, Connaughton, Briggs, Burgess, Ryde,
  Daigne, M{\'e}sz{\'a}ros, Goldstein, McEnery, Omodei,
  {et~al.}}]{guiriec2011detection}
Guiriec, S., Connaughton, V., Briggs, M.~S., {et~al.} 2011, The Astrophysical
  Journal Letters, 727, L33

\bibitem[{Guiriec {et~al.}(2013)Guiriec, Daigne, Hasco{\"e}t, Vianello, Ryde,
  Mochkovitch, Kouveliotou, Xiong, Bhat, Foley, {et~al.}}]{guiriec2013evidence}
Guiriec, S., Daigne, F., Hasco{\"e}t, R., {et~al.} 2013, The Astrophysical
  Journal, 770, 32

\bibitem[{Harris {et~al.}(2020)Harris, Millman, van~der Walt, Gommers,
  Virtanen, Cournapeau, Wieser, Taylor, Berg, Smith, Kern, Picus, Hoyer, van
  Kerkwijk, Brett, Haldane, del R{\'{i}}o, Wiebe, Peterson,
  G{\'{e}}rard-Marchant, Sheppard, Reddy, Weckesser, Abbasi, Gohlke, \&
  Oliphant}]{harris2020array}
Harris, C.~R., Millman, K.~J., van~der Walt, S.~J., {et~al.} 2020, Nature, 585,
  357

\bibitem[{Higson {et~al.}(2019)Higson, Handley, Hobson, \&
  Lasenby}]{higson2019dynamic}
Higson, E., Handley, W., Hobson, M., \& Lasenby, A. 2019, Statistics and
  Computing, 29, 891

\bibitem[{Hou {et~al.}(2018)Hou, Zhang, Meng, Wu, Liang, L{\"u}, Liu, Liang,
  Lin, Lu, {et~al.}}]{hou2018multicolor}
Hou, S.-J., Zhang, B.-B., Meng, Y.-Z., {et~al.} 2018, The Astrophysical
  Journal, 866, 13

\bibitem[{Hunter(2007)}]{Hunter}
Hunter, J.~D. 2007, Computing in Science \& Engineering, 9, 90

\bibitem[{{Jin} {et~al.}(2020){Jin}, {Covino}, {Liao}, {Li}, {D'Avanzo}, {Fan},
  \& {Wei}}]{2020NatAs...4...77J}
{Jin}, Z.-P., {Covino}, S., {Liao}, N.-H., {et~al.} 2020, Nature Astronomy, 4,
  77

\bibitem[{{Jin} {et~al.}(2015){Jin}, {Li}, {Cano}, {Covino}, {Fan}, \&
  {Wei}}]{Jin2015}
{Jin}, Z.-P., {Li}, X., {Cano}, Z., {et~al.} 2015, \apjl, 811, L22

\bibitem[{{Jin} {et~al.}(2016){Jin}, {Hotokezaka}, {Li}, {Tanaka}, {D'Avanzo},
  {Fan}, {Covino}, {Wei}, \& {Piran}}]{2016NatCo...712898J}
{Jin}, Z.-P., {Hotokezaka}, K., {Li}, X., {et~al.} 2016, Nature Communications,
  7, 12898

\bibitem[{Jin {et~al.}(2023)Jin, Zhou, Wang, Geng, Covino, Wu, Li, Fan, Wei, \&
  Wei}]{jin2023detection}
Jin, Z.-P., Zhou, H., Wang, Y., {et~al.} 2023, arXiv preprint arXiv:2301.02407

\bibitem[{Kaneko {et~al.}(2006)Kaneko, Preece, Briggs, Paciesas, Meegan, \&
  Band}]{kaneko2006complete}
Kaneko, Y., Preece, R.~D., Briggs, M.~S., {et~al.} 2006, The Astrophysical
  Journal Supplement Series, 166, 298

\bibitem[{Kobayashi {et~al.}(1997)Kobayashi, Piran,
  {et~al.}}]{kobayashi1997can}
Kobayashi, S., Piran, T., {et~al.} 1997, The Astrophysical Journal, 490, 92

\bibitem[{Kumar(1999)}]{kumar1999gamma}
Kumar, P. 1999, The Astrophysical Journal, 523, L113

\bibitem[{Lazzati {et~al.}(1999)Lazzati, Ghisellini, \&
  Celotti}]{lazzati1999constraints}
Lazzati, D., Ghisellini, G., \& Celotti, A. 1999, Monthly Notices of the Royal
  Astronomical Society, 309, L13

\bibitem[{{Leloudas} {et~al.}(2016){Leloudas}, {Fraser}, {Stone}, {van Velzen},
  {Jonker}, {Arcavi}, {Fremling}, {Maund}, {Smartt}, {Kr{\`\i}hler},
  {Miller-Jones}, {Vreeswijk}, {Gal-Yam}, {Mazzali}, {De Cia}, {Howell},
  {Inserra}, {Patat}, {de Ugarte Postigo}, {Yaron}, {Ashall}, {Bar},
  {Campbell}, {Chen}, {Childress}, {Elias-Rosa}, {Harmanen}, {Hosseinzadeh},
  {Johansson}, {Kangas}, {Kankare}, {Kim}, {Kuncarayakti}, {Lyman}, {Magee},
  {Maguire}, {Malesani}, {Mattila}, {McCully}, {Nicholl}, {Prentice},
  {Romero-Ca{\~n}izales}, {Schulze}, {Smith}, {Sollerman}, {Sullivan},
  {Tucker}, {Valenti}, {Wheeler}, \& {Young}}]{2016NatAs...1E...2L}
{Leloudas}, G., {Fraser}, M., {Stone}, N.~C., {et~al.} 2016, Nature Astronomy,
  1, 0002

\bibitem[{Levan {et~al.}(2023)Levan, Watson, Hjorth, Tanvir, Malesani, Burns,
  Schneider, Fynbo, Vergani, Fong, {et~al.}}]{levan2023grb}
Levan, A., Watson, D., Hjorth, J., {et~al.} 2023, GRB Coordinates Network,
  33580, 1

\bibitem[{Li {et~al.}(2019)Li, Geng, Meng, Wu, Huang, Wang, Moradi, Uhm, \&
  Zhang}]{li2019double}
Li, L., Geng, J.-J., Meng, Y.-Z., {et~al.} 2019, The Astrophysical Journal,
  884, 109

\bibitem[{Li \& Paczy{\'n}ski(1998)}]{li1998transient}
Li, L.-X., \& Paczy{\'n}ski, B. 1998, The Astrophysical Journal, 507, L59

\bibitem[{Liang {et~al.}(2004)Liang, Dai, \& Wu}]{liang2004luminosity}
Liang, E., Dai, Z., \& Wu, X. 2004, The Astrophysical Journal, 606, L29

\bibitem[{Lloyd \& Petrosian(2000)}]{lloyd2000synchrotron}
Lloyd, N.~M., \& Petrosian, V. 2000, The Astrophysical Journal, 543, 722

\bibitem[{Lu {et~al.}(2012)Lu, Wei, Liang, Zhang, L{\"u}, L{\"u}, Lei, \&
  Zhang}]{lu2012comprehensive}
Lu, R.-J., Wei, J.-J., Liang, E.-W., {et~al.} 2012, The Astrophysical Journal,
  756, 112

\bibitem[{Lundman {et~al.}(2013)Lundman, Pe'er, \& Ryde}]{lundman2013theory}
Lundman, C., Pe'er, A., \& Ryde, F. 2013, Monthly Notices of the Royal
  Astronomical Society, 428, 2430

\bibitem[{Maxham \& Zhang(2009)}]{maxham2009modeling}
Maxham, A., \& Zhang, B. 2009, The Astrophysical Journal, 707, 1623

\bibitem[{Meegan {et~al.}(2009)Meegan, Lichti, Bhat, Bissaldi, Briggs,
  Connaughton, Diehl, Fishman, Greiner, Hoover, {et~al.}}]{meegan2009fermi}
Meegan, C., Lichti, G., Bhat, P., {et~al.} 2009, The Astrophysical Journal,
  702, 791

\bibitem[{{Meegan} {et~al.}(2009){Meegan}, {Lichti}, {Bhat}, {Bissaldi},
  {Briggs}, {Connaughton}, {Diehl}, {Fishman}, {Greiner}, {Hoover}, {van der
  Horst}, {von Kienlin}, {Kippen}, {Kouveliotou}, {McBreen}, {Paciesas},
  {Preece}, {Steinle}, {Wallace}, {Wilson}, \&
  {Wilson-Hodge}}]{2009ApJ...702..791M}
{Meegan}, C., {Lichti}, G., {Bhat}, P.~N., {et~al.} 2009, \apj, 702, 791

\bibitem[{Minaev \& Pozanenko(2020{\natexlab{a}})}]{minaev2020p}
Minaev, P.~Y., \& Pozanenko, A. 2020{\natexlab{a}}, Monthly Notices of the
  Royal Astronomical Society, 492, 1919

\bibitem[{Minaev \& Pozanenko(2020{\natexlab{b}})}]{minaev2020grb}
---. 2020{\natexlab{b}}, Astronomy Letters, 46, 573

\bibitem[{Mochkovitch {et~al.}(1995)Mochkovitch, Maitia, \&
  Marques}]{mochkovitch1995internal}
Mochkovitch, R., Maitia, V., \& Marques, R. 1995, Astrophysics and Space
  Science, 231, 441

\bibitem[{Norris {et~al.}(2000)Norris, Marani, \&
  Bonnell}]{norris2000connection}
Norris, J., Marani, G., \& Bonnell, J. 2000, The Astrophysical Journal, 534,
  248

\bibitem[{Norris(2002)}]{norris2002implications}
Norris, J.~P. 2002, The Astrophysical Journal, 579, 386

\bibitem[{Norris {et~al.}(2005)Norris, Bonnell, Kazanas, Scargle, Hakkila, \&
  Giblin}]{norris2005long}
Norris, J.~P., Bonnell, J.~T., Kazanas, D., {et~al.} 2005, The Astrophysical
  Journal, 627, 324

\bibitem[{Paczynski(1986)}]{paczynski1986gamma}
Paczynski, B. 1986, The Astrophysical Journal, 308, L43

\bibitem[{Panaitescu {et~al.}(1999)Panaitescu, Spada, \&
  M{\'e}sz{\'a}ros}]{panaitescu1999power}
Panaitescu, A., Spada, M., \& M{\'e}sz{\'a}ros, P. 1999, The Astrophysical
  Journal, 522, L105

\bibitem[{Pe’er(2008)}]{pe2008temporal}
Pe’er, A. 2008, The Astrophysical Journal, 682, 463

\bibitem[{Pe’er \& Ryde(2017)}]{pe2017photospheric}
Pe’er, A., \& Ryde, F. 2017, International Journal of Modern Physics D, 26,
  1730018

\bibitem[{Preece {et~al.}(2002)Preece, Briggs, Giblin, Mallozzi, Pendleton,
  Paciesas, \& Band}]{preece2002consistency}
Preece, R., Briggs, M., Giblin, T., {et~al.} 2002, The Astrophysical Journal,
  581, 1248

\bibitem[{Preece {et~al.}(2014)Preece, Burgess, Von~Kienlin, Bhat, Briggs,
  Byrne, Chaplin, Cleveland, Collazzi, Connaughton, {et~al.}}]{preece2014first}
Preece, R., Burgess, J.~M., Von~Kienlin, A., {et~al.} 2014, Science, 343, 51

\bibitem[{Preece {et~al.}(2000)Preece, Briggs, Mallozzi, Pendleton, Paciesas,
  \& Band}]{preece2000batse}
Preece, R.~D., Briggs, M.~S., Mallozzi, R.~S., {et~al.} 2000, The Astrophysical
  Journal Supplement Series, 126, 19

\bibitem[{Preece {et~al.}(1998)Preece, Briggs, Mallozzi, Pendleton, Paciesas,
  \& Band}]{preece1998synchrotron}
---. 1998, The Astrophysical Journal, 506, L23

\bibitem[{Ren {et~al.}(2022)Ren, Wang, \& Zhang}]{ren2022very}
Ren, J., Wang, Y., \& Zhang, L.-L. 2022, arXiv preprint arXiv:2210.10673

\bibitem[{{Ryde} {et~al.}(2017){Ryde}, {Lundman}, \&
  {Acuner}}]{Ryde2017Emission}
{Ryde}, F., {Lundman}, C., \& {Acuner}, Z. 2017, \mnras, 472, 1897

\bibitem[{{Ryde} {et~al.}(2019){Ryde}, {Yu}, {Dereli-B{\'e}gu{\'e}}, {Lundman},
  {Pe'er}, \& {Li}}]{Ryde2019intensity}
{Ryde}, F., {Yu}, H.-F., {Dereli-B{\'e}gu{\'e}}, H., {et~al.} 2019, \mnras,
  484, 1912

\bibitem[{Ryde {et~al.}(2010)Ryde, Axelsson, Zhang, McGlynn, Pe'er, Lundman,
  Larsson, Battelino, Zhang, Bissaldi, {et~al.}}]{ryde2010identification}
Ryde, F., Axelsson, M., Zhang, B., {et~al.} 2010, The Astrophysical Journal
  Letters, 709, L172

\bibitem[{Scargle {et~al.}(2013)Scargle, Norris, Jackson, \&
  Chiang}]{scargle2013studies}
Scargle, J.~D., Norris, J.~P., Jackson, B., \& Chiang, J. 2013, The
  Astrophysical Journal, 764, 167

\bibitem[{Skilling(2006)}]{skilling2006nested}
Skilling, J. 2006, Bayesian analysis, 1, 833

\bibitem[{Spada {et~al.}(2000)Spada, Panaitescu, \&
  Meszaros}]{spada2000analysis}
Spada, M., Panaitescu, A., \& Meszaros, P. 2000, The Astrophysical Journal,
  537, 824

\bibitem[{Speagle(2020)}]{speagle2020dynesty}
Speagle, J.~S. 2020, Monthly Notices of the Royal Astronomical Society, 493,
  3132

\bibitem[{Svinkin {et~al.}(2023c)Svinkin, Frederiks, Ridnaia, Tsvetkova,
  Lysenko, Team, {et~al.}}]{svinkin2023grb}
Svinkin, D., Frederiks, D., Ridnaia, A., {et~al.} 2023c, GRB Coordinates
  Network, 33579, 1

\bibitem[{{Svinkin} {et~al.}(2023{\natexlab{a}}){Svinkin}, {Frederiks},
  {Ulanov}, {Tsvetkova}, {Lysenko}, {Ridnaia}, {Cline}, \& {Konus-Wind
  Team}}]{2023GCN.33427....1S}
{Svinkin}, D., {Frederiks}, D., {Ulanov}, M., {et~al.} 2023{\natexlab{a}}, GRB
  Coordinates Network, 33427, 1

\bibitem[{{Svinkin} {et~al.}(2023{\natexlab{b}}){Svinkin}, {Frederiks},
  {Ulanov}, {Tsvetkova}, {Lysenko}, {Ridnaia}, {Cline}, \& team}]{GCN33427}
---. 2023{\natexlab{b}}, GRB Coordinates Network, 33414, 1

\bibitem[{Tavani(1996)}]{tavani1996shock}
Tavani, M. 1996, The Astrophysical Journal, 466, 768

\bibitem[{{the GRBAlpha collaboration}.(2023)}]{GCN33418}
{the GRBAlpha collaboration}. 2023, GRB Coordinates Network, 334148, 1

\bibitem[{Thrane \& Talbot(2019)}]{thrane2019introduction}
Thrane, E., \& Talbot, C. 2019, Publications of the Astronomical Society of
  Australia, 36

\bibitem[{Uhm \& Zhang(2014)}]{uhm2014fast}
Uhm, Z.~L., \& Zhang, B. 2014, Nature Physics, 10, 351

\bibitem[{Ukwatta {et~al.}(2010)Ukwatta, Stamatikos, Dhuga, Sakamoto,
  Barthelmy, Eskandarian, Gehrels, Maximon, Norris, \&
  Parke}]{ukwatta2010spectral}
Ukwatta, T., Stamatikos, M., Dhuga, K., {et~al.} 2010, The Astrophysical
  Journal, 711, 1073

\bibitem[{van~de Schoot {et~al.}(2021)van~de Schoot, Depaoli, King, Kramer,
  M{\"a}rtens, Tadesse, Vannucci, Gelman, Veen, Willemsen,
  {et~al.}}]{van2021bayesian}
van~de Schoot, R., Depaoli, S., King, R., {et~al.} 2021, Nature Reviews Methods
  Primers, 1, 1

\bibitem[{{Vurm} \& {Beloborodov}(2016)}]{Vurm2016Radiative}
{Vurm}, I., \& {Beloborodov}, A.~M. 2016, \apj, 831, 175

\bibitem[{Wang {et~al.}(2022)Wang, Zheng, \& Jin}]{wang2022grb}
Wang, Y., Zheng, T.-C., \& Jin, Z.-P. 2022, The Astrophysical Journal, 940, 142

\bibitem[{Wei \& Gao(2003)}]{wei2003there}
Wei, D., \& Gao, W. 2003, Monthly Notices of the Royal Astronomical Society,
  345, 743

\bibitem[{{Xia} {et~al.}(2022){Xia}, {Wang}, {Yuan}, \&
  {Fan}}]{2022arXiv221013052X}
{Xia}, Z.-Q., {Wang}, Y., {Yuan}, Q., \& {Fan}, Y.-Z. 2022, arXiv e-prints,
  arXiv:2210.13052

\bibitem[{{Xiong} {et~al.}(2023){Xiong}, {Wang}, {Huang}, \& team}]{GCN33406}
{Xiong}, S.-L., {Wang}, C.-W., {Huang}, Y., \& team, G. 2023, GRB Coordinates
  Network, 33406, 1

\bibitem[{{Yang} {et~al.}(2015){Yang}, {Jin}, {Li}, {Covino}, {Zheng},
  {Hotokezaka}, {Fan}, {Piran}, \& {Wei}}]{Yang2015}
{Yang}, B., {Jin}, Z.-P., {Li}, X., {et~al.} 2015, Nature Communications, 6,
  7323

\bibitem[{Yang {et~al.}(2022)Yang, Ai, Zhang, Zhang, Liu, Wang, Yang, Yin, Li,
  \& L{\"u}}]{yang2022long}
Yang, J., Ai, S., Zhang, B.-B., {et~al.} 2022, Nature, 612, 232

\bibitem[{Yang {et~al.}(2023)Yang, Zhao, Yan, Wang, Zhang, An, Cai, Li, Li,
  Liu, {et~al.}}]{yang2023synchrotron}
Yang, J., Zhao, X.-H., Yan, Z., {et~al.} 2023, arXiv preprint arXiv:2303.00898

\bibitem[{Yonetoku {et~al.}(2004)Yonetoku, Murakami, Nakamura, Yamazaki, Inoue,
  \& Ioka}]{yonetoku2004gamma}
Yonetoku, D., Murakami, T., Nakamura, T., {et~al.} 2004, The Astrophysical
  Journal, 609, 935

\bibitem[{{Yu} {et~al.}(2019){Yu}, {Dereli-B{\'e}gu{\'e}}, \&
  {Ryde}}]{yu2019Bayesian}
{Yu}, H.-F., {Dereli-B{\'e}gu{\'e}}, H., \& {Ryde}, F. 2019, \apj, 886, 20

\bibitem[{Zhang \& M{\'e}sz{\'a}ros(2002)}]{zhang2002analysis}
Zhang, B., \& M{\'e}sz{\'a}ros, P. 2002, The Astrophysical Journal, 581, 1236

\bibitem[{{Zhang} \& {Yan}(2011)}]{zhang2010internal}
{Zhang}, B., \& {Yan}, H. 2011, \apj, 726, 90

\bibitem[{Zhang {et~al.}(2018)Zhang, Zhang, Castro-Tirado, Dai, Tam, Wang, Hu,
  Karpov, Pozanenko, Zhang, {et~al.}}]{zhang2018transition}
Zhang, B.-B., Zhang, B., Castro-Tirado, A.~J., {et~al.} 2018, Nature Astronomy,
  2, 69

\end{thebibliography}
\appendix
\section{HEtools}
\begin{figure}[!h]
	\centering
	\includegraphics[width=0.8\textwidth]{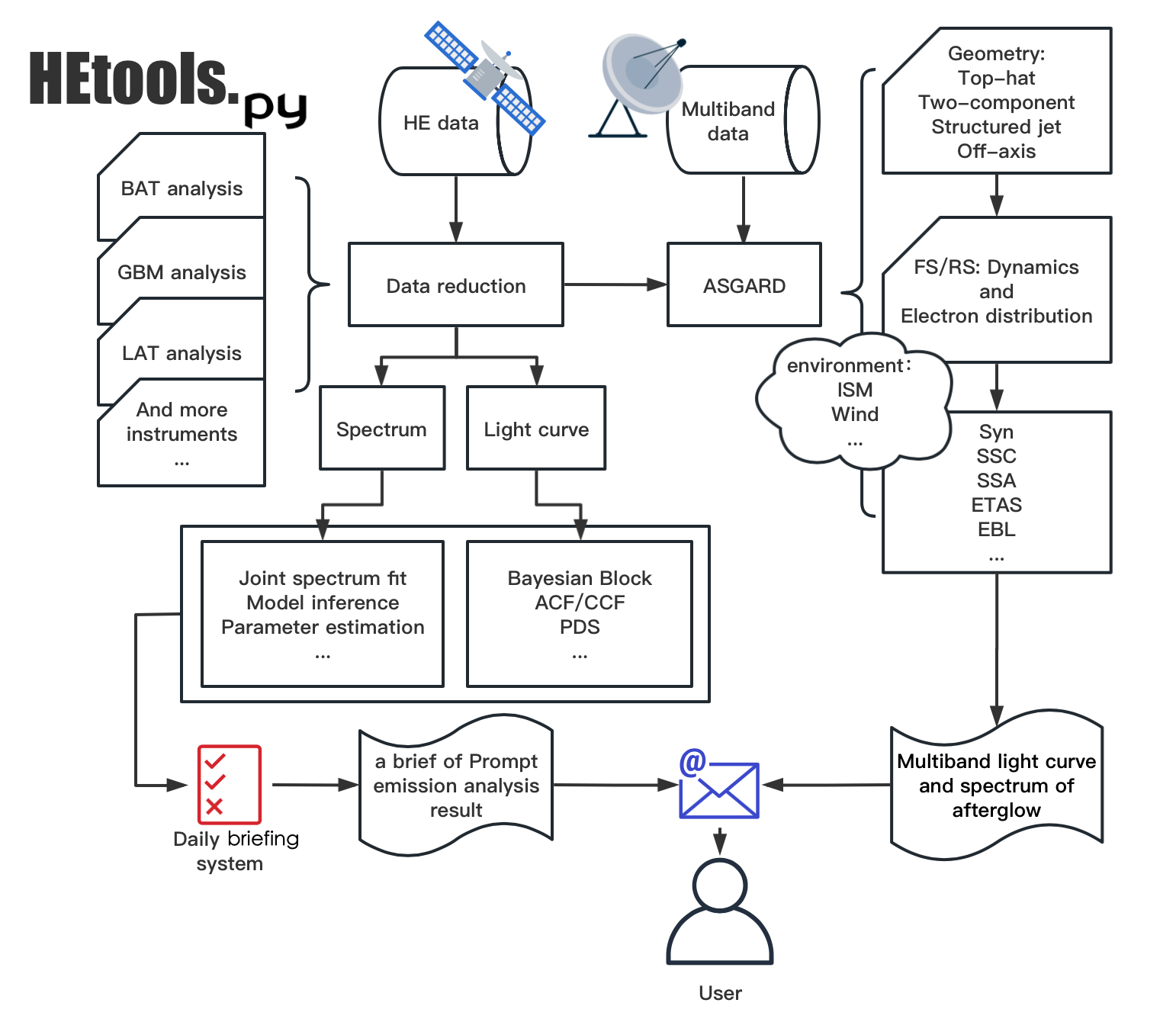}
	\caption{Framework of {\tt HEtools}.}
	\label{fig:HEtools}
\end{figure}

The development of high-energy detection data analysis tools ({\tt HEtools}) is the preparation for the future Very Large Area gamma-ray Space Telescope \citep[VLAST;][]{fan2022very}.
Its original intention is to provide reasonable observational references in order to optimize and expand the scientific output of VLAST, e.g. to make reasonable observational arrangements for the cascading radiation of GRB veay high energy (VHE) photons under the influence of intergalactic magnetic field \citep{2022arXiv221013052X}.
In order to test the feasibility of this tool, we apply it to the study of GRB.
As shown in Figure \ref{fig:HEtools}, {\tt HEtools} has separate modules for data analysis of different instruments, currently including but not limited to Swift-BAT, Fermi-GBM/LAT.
Python-based frameworks are very friendly to future instrument support, such as VLAST.
The corresponding light curve and energy spectrum can be generated through online data retrieval, and the preset spectrum inference or time series analysis can be performed, and finally a briefing is generated and sent to the user's mailbox.
Based on friendly expansion capabilities, by combining with A Standard Gamma-ray burst Afterglow Radiation Diagnoser \citep[{\tt ASGARD};][]{ren2022very}, it is used to {model} afterglow light curve and spectrum by considering various situations, various environments, and various physical processes.
It is possible to constrain the microscopic physical parameters of GRB by using a sampling method (e.g. {\tt MCMC}, {\tt Nested}, etc) for Bayesian inference.
And more features are under continuous development.

\label{app:1}

\section{Spectral inference result}
\begin{longtable}{ccccccccc} 
\label{tab:1}\\
\caption{Spectral inference result}\\
\toprule
{Time Interval} & & {CPL Model} & & & {Band Model} & & & {Higher Evidence}\\
{[s]} & $\alpha$ & $E_{\rm p}$ [keV]& $\ln{\cal Z}$  & $\alpha$ & $\beta$ & $E_{\rm p}$ [keV] &  $\ln{\cal Z}$& \\
\endfirsthead
\caption{continued}\\
\toprule
{Time Interval} & & {CPL Model} & & & {Band Model} & & & {Higher Evidence}\\
{[s]} & $\alpha$ & $E_{\rm p}$ [keV]& $\ln{\cal Z}$  & $\alpha$ & $\beta$ & $E_{\rm p}$ [keV] &  $\ln{\cal Z}$& \\
\hline
\endhead
\hline
\endlastfoot
\hline
{[-0.02 , 0.75]} & -0.989 $\pm$ 0.019 & 205.66 $\pm$ 3.41 & -348.83 & -0.884 $\pm$ 0.028 & -2.93 $\pm$ 0.09 & 180.66 $\pm$ 4.87 & -333.77 & Band \\ 
{[0.75 , 0.88]} & -0.495 $\pm$ 0.021 & 646.15 $\pm$ 12.81 & -227.72 & -0.485 $\pm$ 0.022 & -4.33 $\pm$ 1.49 & 636.80 $\pm$ 13.28 & -228.82 & CPL \\ 
{[0.88 , 0.94]} & -0.608 $\pm$ 0.035 & 633.67 $\pm$ 24.19 & -197.83 & -0.613 $\pm$ 0.034 & -8.89 $\pm$ 1.75 & 637.09 $\pm$ 23.71 & -200.55 & CPL \\ 
{[0.94 , 1.07]} & -0.410 $\pm$ 0.022 & 646.51 $\pm$ 11.65 & -216.47 & -0.397 $\pm$ 0.022 & -4.68 $\pm$ 1.58 & 637.84 $\pm$ 11.72 & -217.46 & CPL \\ 
{[1.07 , 1.46]} & -0.601 $\pm$ 0.014 & 534.75 $\pm$ 7.34 & -249.33 & -0.597 $\pm$ 0.014 & -5.14 $\pm$ 1.62 & 532.07 $\pm$ 7.30 & -251.61 & CPL \\ 
{[1.46 , 1.52]} & -0.427 $\pm$ 0.022 & 1179.50 $\pm$ 25.82 & -240.47 & -0.404 $\pm$ 0.025 & -3.92 $\pm$ 0.72 & 1140.25 $\pm$ 28.56 & -239.12 & Band \\ 
{[1.52 , 1.65]} & -0.269 $\pm$ 0.016 & 992.74 $\pm$ 11.49 & -308.71 & -0.269 $\pm$ 0.015 & -7.38 $\pm$ 1.21 & 993.53 $\pm$ 11.33 & -310.53 & CPL \\ 
{[1.65 , 1.84]} & -0.353 $\pm$ 0.012 & 1339.53 $\pm$ 13.87 & -319.24 & -0.353 $\pm$ 0.012 & -6.02 $\pm$ 1.23 & 1340.09 $\pm$ 14.22 & -320.75 & CPL \\ 
{[1.84 , 1.97]} & -0.238 $\pm$ 0.015 & 1216.99 $\pm$ 13.41 & -296.16 & -0.228 $\pm$ 0.016 & -5.07 $\pm$ 0.92 & 1205.26 $\pm$ 14.47 & -295.39 & Band \\ 
{[1.97 , 2.29]} & -0.250 $\pm$ 0.012 & 814.15 $\pm$ 6.74 & -392.14 & -0.245 $\pm$ 0.012 & -5.46 $\pm$ 1.07 & 811.38 $\pm$ 7.22 & -391.65 & Band \\ 
{[2.29 , 2.67]} & -0.258 $\pm$ 0.012 & 722.73 $\pm$ 6.40 & -266.76 & -0.258 $\pm$ 0.012 & -9.97 $\pm$ 1.19 & 722.41 $\pm$ 6.34 & -268.30 & CPL \\ 
{[Bad time]} &... &... & ... &... & ... & ... & ... & ... \\
{[7.22 , 7.28]} & -0.583 $\pm$ 0.020 & 898.18 $\pm$ 20.53 & -204.75 & -0.567 $\pm$ 0.021 & -4.01 $\pm$ 1.09 & 874.17 $\pm$ 23.02 & -204.98 & CPL \\ 
{[7.28 , 7.47]} & -0.718 $\pm$ 0.013 & 755.90 $\pm$ 12.06 & -211.25 & -0.710 $\pm$ 0.014 & -4.27 $\pm$ 1.38 & 742.98 $\pm$ 13.37 & -212.98 & CPL \\ 
{[7.47 , 7.66]} & -0.789 $\pm$ 0.016 & 516.51 $\pm$ 10.44 & -232.59 & -0.761 $\pm$ 0.021 & -3.59 $\pm$ 1.68 & 492.78 $\pm$ 14.99 & -234.84 & CPL \\ 
{[7.66 , 7.98]} & -0.762 $\pm$ 0.009 & 742.30 $\pm$ 9.09 & -294.85 & -0.760 $\pm$ 0.009 & -5.16 $\pm$ 1.46 & 738.18 $\pm$ 9.50 & -297.19 & CPL \\ 
{[7.98 , 8.62]} & -0.682 $\pm$ 0.005 & 1208.32 $\pm$ 8.13 & -416.09 & -0.677 $\pm$ 0.005 & -4.95 $\pm$ 0.46 & 1198.17 $\pm$ 9.22 & -414.64 & Band \\ 
{[8.62 , 8.88]} & -0.681 $\pm$ 0.010 & 888.19 $\pm$ 11.28 & -274.44 & -0.679 $\pm$ 0.010 & -5.25 $\pm$ 1.46 & 883.35 $\pm$ 11.45 & -276.92 & CPL \\ 
{[8.88 , 9.33]} & -0.741 $\pm$ 0.008 & 819.98 $\pm$ 8.74 & -291.80 & -0.735 $\pm$ 0.009 & -4.65 $\pm$ 1.21 & 811.19 $\pm$ 10.20 & -294.20 & CPL \\ 
{[9.33 , 9.52]} & -0.708 $\pm$ 0.011 & 783.93 $\pm$ 11.14 & -267.81 & -0.703 $\pm$ 0.012 & -4.50 $\pm$ 1.66 & 775.44 $\pm$ 12.79 & -270.09 & CPL \\ 
{[9.52 , 9.71]} & -0.789 $\pm$ 0.011 & 1045.08 $\pm$ 17.27 & -235.41 & -0.781 $\pm$ 0.012 & -4.31 $\pm$ 1.87 & 1028.52 $\pm$ 19.43 & -239.10 & CPL \\ 
{[9.71 , 9.97]} & -0.768 $\pm$ 0.010 & 1062.86 $\pm$ 16.12 & -275.58 & -0.766 $\pm$ 0.010 & -4.38 $\pm$ 1.57 & 1051.19 $\pm$ 17.40 & -278.17 & CPL \\ 
{[9.97 , 10.10]} & -0.743 $\pm$ 0.013 & 1131.57 $\pm$ 21.65 & -262.97 & -0.743 $\pm$ 0.013 & -5.82 $\pm$ 1.63 & 1127.88 $\pm$ 21.93 & -266.25 & CPL \\ 
{[10.10 , 10.35]} & -0.711 $\pm$ 0.009 & 1120.10 $\pm$ 13.90 & -281.95 & -0.699 $\pm$ 0.009 & -4.06 $\pm$ 0.24 & 1092.27 $\pm$ 15.49 & -275.88 & Band \\ 
{[10.35 , 10.54]} & -0.720 $\pm$ 0.013 & 850.01 $\pm$ 14.60 & -248.45 & -0.712 $\pm$ 0.014 & -4.07 $\pm$ 1.37 & 831.56 $\pm$ 16.84 & -250.31 & CPL \\ 
{[10.54 , 10.67]} & -0.759 $\pm$ 0.012 & 975.23 $\pm$ 17.78 & -224.87 & -0.758 $\pm$ 0.013 & -4.40 $\pm$ 1.78 & 967.61 $\pm$ 19.52 & -228.18 & CPL \\ 
{[10.67 , 10.99]} & -0.796 $\pm$ 0.010 & 820.59 $\pm$ 11.60 & -250.28 & -0.795 $\pm$ 0.010 & -5.49 $\pm$ 1.55 & 817.07 $\pm$ 11.40 & -254.16 & CPL \\ 
{[10.99 , 11.18]} & -0.841 $\pm$ 0.015 & 673.09 $\pm$ 14.68 & -268.75 & -0.837 $\pm$ 0.015 & -4.31 $\pm$ 1.86 & 669.09 $\pm$ 15.55 & -272.80 & CPL \\ 
{[11.18 , 11.44]} & -0.837 $\pm$ 0.010 & 841.01 $\pm$ 12.79 & -282.62 & -0.836 $\pm$ 0.010 & -4.99 $\pm$ 1.59 & 837.96 $\pm$ 13.08 & -286.08 & CPL \\ 
{[11.44 , 11.76]} & -0.894 $\pm$ 0.012 & 599.32 $\pm$ 10.22 & -300.34 & -0.888 $\pm$ 0.012 & -4.18 $\pm$ 1.62 & 590.97 $\pm$ 11.69 & -303.59 & CPL \\ 
{[11.76 , 11.82]} & -0.980 $\pm$ 0.032 & 513.66 $\pm$ 26.26 & -186.13 & -0.982 $\pm$ 0.032 & -9.22 $\pm$ 1.76 & 516.04 $\pm$ 25.98 & -190.74 & CPL \\ 
{[11.82 , 12.08]} & -0.891 $\pm$ 0.009 & 944.59 $\pm$ 15.19 & -317.53 & -0.889 $\pm$ 0.009 & -5.16 $\pm$ 1.67 & 942.72 $\pm$ 15.91 & -321.37 & CPL \\ 
{[12.08 , 12.66]} & -0.935 $\pm$ 0.007 & 907.54 $\pm$ 11.30 & -320.10 & -0.933 $\pm$ 0.007 & -5.02 $\pm$ 1.62 & 904.34 $\pm$ 11.32 & -324.15 & CPL \\ 
{[12.66 , 12.91]} & -0.870 $\pm$ 0.011 & 759.83 $\pm$ 12.37 & -277.93 & -0.862 $\pm$ 0.012 & -3.94 $\pm$ 1.25 & 744.56 $\pm$ 15.95 & -279.45 & CPL \\ 
{[12.91 , 13.04]} & -0.910 $\pm$ 0.012 & 780.91 $\pm$ 16.11 & -276.48 & -0.888 $\pm$ 0.014 & -3.62 $\pm$ 0.34 & 740.75 $\pm$ 19.34 & -274.54 & Band \\ 
{[13.04 , 13.36]} & -0.972 $\pm$ 0.008 & 868.19 $\pm$ 13.29 & -294.04 & -0.969 $\pm$ 0.008 & -4.54 $\pm$ 1.50 & 862.00 $\pm$ 13.89 & -297.40 & CPL \\ 
{[13.36 , 13.49]} & -1.014 $\pm$ 0.014 & 859.05 $\pm$ 23.65 & -223.34 & -1.006 $\pm$ 0.014 & -4.24 $\pm$ 1.89 & 844.43 $\pm$ 24.92 & -227.54 & CPL \\ 
{[13.49 , 14.00]} & -1.053 $\pm$ 0.009 & 740.81 $\pm$ 12.80 & -353.83 & -1.048 $\pm$ 0.009 & -4.19 $\pm$ 1.85 & 730.31 $\pm$ 14.15 & -357.77 & CPL \\ 
{[14.00 , 14.38]} & -1.045 $\pm$ 0.008 & 902.13 $\pm$ 15.62 & -324.21 & -1.041 $\pm$ 0.009 & -4.05 $\pm$ 1.29 & 886.72 $\pm$ 18.23 & -327.35 & CPL \\ 
{[14.38 , 14.51]} & -1.089 $\pm$ 0.022 & 511.97 $\pm$ 21.33 & -237.81 & -1.071 $\pm$ 0.026 & -3.18 $\pm$ 2.23 & 485.06 $\pm$ 26.48 & -242.08 & CPL \\ 
{[14.51 , 14.96]} & -1.161 $\pm$ 0.010 & 595.94 $\pm$ 12.24 & -362.61 & -1.151 $\pm$ 0.011 & -3.66 $\pm$ 1.64 & 578.34 $\pm$ 14.68 & -366.03 & CPL \\ 
{[14.96 , 15.15]} & -1.042 $\pm$ 0.012 & 718.04 $\pm$ 17.20 & -262.66 & -1.037 $\pm$ 0.012 & -4.05 $\pm$ 1.95 & 707.72 $\pm$ 17.91 & -266.76 & CPL \\ 
{[15.15 , 15.79]} & -1.100 $\pm$ 0.008 & 610.22 $\pm$ 9.94 & -387.46 & -1.098 $\pm$ 0.008 & -4.67 $\pm$ 1.72 & 608.71 $\pm$ 9.94 & -391.12 & CPL \\ 
{[15.79 , 16.69]} & -1.110 $\pm$ 0.007 & 773.09 $\pm$ 11.78 & -418.04 & -1.104 $\pm$ 0.007 & -3.88 $\pm$ 0.67 & 757.67 $\pm$ 13.70 & -419.30 & CPL \\ 
{[16.69 , 17.46]} & -1.177 $\pm$ 0.009 & 542.63 $\pm$ 10.64 & -333.94 & -1.178 $\pm$ 0.009 & -9.52 $\pm$ 1.57 & 541.82 $\pm$ 10.65 & -339.17 & CPL \\ 
{[17.46 , 17.65]} & -1.131 $\pm$ 0.023 & 437.91 $\pm$ 17.85 & -225.09 & -1.134 $\pm$ 0.024 & -4.07 $\pm$ 1.99 & 436.03 $\pm$ 17.93 & -230.21 & CPL \\ 
{[17.65 , 17.90]} & -1.207 $\pm$ 0.033 & 255.46 $\pm$ 11.56 & -232.97 & -1.193 $\pm$ 0.035 & -2.97 $\pm$ 1.62 & 244.57 $\pm$ 12.66 & -236.99 & CPL \\ 
{[17.90 , 18.67]} & -1.510 $\pm$ 0.026 & 211.49 $\pm$ 10.72 & -246.95 & -1.499 $\pm$ 0.027 & -3.04 $\pm$ 2.26 & 204.69 $\pm$ 10.93 & -253.03 & CPL \\ 
{[18.67 , 18.99]} & -1.383 $\pm$ 0.023 & 361.83 $\pm$ 20.30 & -262.07 & -1.383 $\pm$ 0.023 & -4.82 $\pm$ 1.94 & 364.25 $\pm$ 19.76 & -268.33 & CPL \\ 
{[18.99 , 19.12]} & -1.297 $\pm$ 0.031 & 399.53 $\pm$ 28.47 & -206.72 & -1.173 $\pm$ 0.056 & -2.39 $\pm$ 0.79 & 284.87 $\pm$ 41.62 & -209.25 & CPL \\ 
{[19.12 , 19.82]} & -1.251 $\pm$ 0.009 & 546.93 $\pm$ 12.20 & -374.74 & -1.248 $\pm$ 0.010 & -3.84 $\pm$ 1.98 & 539.93 $\pm$ 12.87 & -379.88 & CPL \\ 
{[19.82 , 20.40]} & -1.215 $\pm$ 0.008 & 808.29 $\pm$ 17.91 & -393.03 & -1.214 $\pm$ 0.008 & -4.34 $\pm$ 1.94 & 799.98 $\pm$ 18.51 & -398.04 & CPL \\ 
{[20.40 , 21.10]} & -1.157 $\pm$ 0.010 & 601.17 $\pm$ 12.79 & -331.86 & -1.157 $\pm$ 0.010 & -5.14 $\pm$ 1.72 & 599.88 $\pm$ 12.97 & -337.31 & CPL \\ 
{[21.10 , 21.68]} & -1.144 $\pm$ 0.009 & 616.20 $\pm$ 12.88 & -343.02 & -1.128 $\pm$ 0.011 & -3.35 $\pm$ 0.21 & 587.33 $\pm$ 14.78 & -339.28 & Band \\ 
{[21.68 , 23.22]} & -1.195 $\pm$ 0.007 & 513.65 $\pm$ 7.77 & -414.16 & -1.195 $\pm$ 0.007 & -7.97 $\pm$ 1.42 & 513.74 $\pm$ 7.62 & -420.12 & CPL \\ 
{[23.22 , 23.54]} & -1.182 $\pm$ 0.026 & 241.79 $\pm$ 7.65 & -209.84 & -1.179 $\pm$ 0.027 & -3.59 $\pm$ 2.12 & 238.17 $\pm$ 7.84 & -215.21 & CPL \\ 
{[23.54 , 25.01]} & -1.285 $\pm$ 0.007 & 523.19 $\pm$ 9.34 & -413.35 & -1.286 $\pm$ 0.007 & -6.52 $\pm$ 1.64 & 523.90 $\pm$ 9.33 & -419.18 & CPL \\ 
{[25.01 , 25.20]} & -1.191 $\pm$ 0.029 & 312.28 $\pm$ 13.61 & -239.49 & -1.185 $\pm$ 0.029 & -4.60 $\pm$ 1.90 & 310.79 $\pm$ 13.34 & -245.16 & CPL \\ 
{[25.20 , 25.65]} & -1.336 $\pm$ 0.014 & 379.73 $\pm$ 10.97 & -288.55 & -1.285 $\pm$ 0.030 & -2.74 $\pm$ 2.26 & 326.93 $\pm$ 27.96 & -294.32 & CPL \\ 
{[25.65 , 25.78]} & -1.454 $\pm$ 0.030 & 332.57 $\pm$ 24.93 & -192.48 & -1.453 $\pm$ 0.031 & -4.43 $\pm$ 2.07 & 331.63 $\pm$ 25.57 & -199.54 & CPL \\ 
{[25.78 , 25.97]} & -1.305 $\pm$ 0.022 & 342.05 $\pm$ 13.91 & -226.09 & -1.304 $\pm$ 0.022 & -7.41 $\pm$ 1.87 & 342.16 $\pm$ 14.04 & -231.78 & CPL \\ 
{[25.97 , 26.16]} & -1.407 $\pm$ 0.030 & 221.10 $\pm$ 10.52 & -222.82 & -1.394 $\pm$ 0.031 & -3.41 $\pm$ 2.13 & 217.30 $\pm$ 11.40 & -228.83 & CPL \\ 
{[26.16 , 26.54]} & -1.460 $\pm$ 0.026 & 157.06 $\pm$ 5.34 & -226.38 & -1.374 $\pm$ 0.052 & -2.70 $\pm$ 2.05 & 133.85 $\pm$ 12.03 & -231.84 & CPL \\ 
{[26.54 , 26.74]} & -1.539 $\pm$ 0.024 & 303.73 $\pm$ 20.87 & -258.59 & -1.502 $\pm$ 0.037 & -2.50 $\pm$ 1.24 & 261.14 $\pm$ 29.33 & -263.53 & CPL \\ 
{[26.74 , 27.95]} & -1.562 $\pm$ 0.011 & 354.38 $\pm$ 12.06 & -390.35 & -1.561 $\pm$ 0.011 & -9.30 $\pm$ 1.73 & 353.78 $\pm$ 11.88 & -398.72 & CPL \\ 
{[27.95 , 28.46]} & -1.434 $\pm$ 0.011 & 568.87 $\pm$ 21.29 & -286.64 & -1.434 $\pm$ 0.011 & -3.93 $\pm$ 1.93 & 569.16 $\pm$ 20.82 & -293.03 & CPL \\ 
{[28.46 , 28.72]} & -1.443 $\pm$ 0.019 & 601.31 $\pm$ 43.14 & -215.41 & -1.434 $\pm$ 0.019 & -3.29 $\pm$ 2.21 & 579.88 $\pm$ 43.22 & -221.65 & CPL \\ 
{[28.72 , 30.19]} & -1.435 $\pm$ 0.010 & 509.42 $\pm$ 16.94 & -330.77 & -1.436 $\pm$ 0.010 & -4.27 $\pm$ 1.95 & 508.38 $\pm$ 17.27 & -337.68 & CPL \\ 
{[30.19 , 31.02]} & -1.368 $\pm$ 0.021 & 300.46 $\pm$ 12.67 & -274.87 & -1.367 $\pm$ 0.021 & -9.62 $\pm$ 1.92 & 299.86 $\pm$ 12.86 & -281.80 & CPL \\ 
{[31.02 , 31.98]} & -1.455 $\pm$ 0.023 & 169.70 $\pm$ 5.69 & -259.46 & -1.457 $\pm$ 0.024 & -5.83 $\pm$ 1.85 & 169.84 $\pm$ 5.76 & -265.57 & CPL \\ 
{[31.98 , 32.82]} & -1.604 $\pm$ 0.029 & 119.93 $\pm$ 4.98 & -265.19 & -1.603 $\pm$ 0.030 & -9.04 $\pm$ 1.84 & 120.21 $\pm$ 5.09 & -273.58 & CPL \\ 
{[32.82 , 34.74]} & -1.526 $\pm$ 0.013 & 261.60 $\pm$ 7.58 & -350.06 & -1.525 $\pm$ 0.013 & -9.74 $\pm$ 1.84 & 262.29 $\pm$ 7.53 & -358.65 & CPL \\ 
{[34.74 , 35.63]} & -1.503 $\pm$ 0.016 & 343.03 $\pm$ 14.96 & -276.88 & -1.500 $\pm$ 0.016 & -3.46 $\pm$ 2.13 & 338.91 $\pm$ 15.19 & -285.07 & CPL \\ 
{[35.63 , 37.49]} & -1.531 $\pm$ 0.015 & 196.96 $\pm$ 5.36 & -302.39 & -1.529 $\pm$ 0.015 & -5.57 $\pm$ 1.86 & 196.63 $\pm$ 5.37 & -310.48 & CPL \\ 
{[37.49 , 39.02]} & -1.426 $\pm$ 0.022 & 193.36 $\pm$ 6.32 & -247.76 & -1.417 $\pm$ 0.022 & -3.25 $\pm$ 2.17 & 189.18 $\pm$ 6.46 & -255.64 & CPL \\ 
{[39.02 , 40.37]} & -1.577 $\pm$ 0.025 & 214.94 $\pm$ 12.43 & -288.73 & -1.576 $\pm$ 0.024 & -7.22 $\pm$ 1.90 & 215.57 $\pm$ 11.82 & -301.05 & CPL \\ 
{[40.37 , 41.65]} & -1.651 $\pm$ 0.042 & 89.60 $\pm$ 5.44 & -241.34 & -1.658 $\pm$ 0.054 & -4.64 $\pm$ 1.94 & 89.33 $\pm$ 5.80 & -255.13 & CPL \\ 
{[41.65 , 44.66]} & -1.745 $\pm$ 0.018 & 137.80 $\pm$ 6.08 & -274.03 & -1.745 $\pm$ 0.018 & -8.25 $\pm$ 1.81 & 138.73 $\pm$ 6.10 & -290.60 & CPL \\ 
{[44.66 , 45.94]} & -1.657 $\pm$ 0.052 & 82.54 $\pm$ 6.50 & -215.06 & -1.659 $\pm$ 0.052 & -3.94 $\pm$ 2.03 & 81.62 $\pm$ 6.22 & -220.89 & CPL \\ 
{[45.94 , 47.98]} & -1.601 $\pm$ 0.021 & 202.42 $\pm$ 10.01 & -270.32 & -1.593 $\pm$ 0.022 & -3.03 $\pm$ 2.19 & 195.88 $\pm$ 10.22 & -285.71 & CPL \\ 
{[47.98 , 51.31]} & -1.506 $\pm$ 0.024 & 151.10 $\pm$ 5.15 & -303.35 & -1.488 $\pm$ 0.025 & -2.97 $\pm$ 2.21 & 145.00 $\pm$ 5.70 & -310.94 & CPL \\ 
{[51.31 , 52.66]} & -1.616 $\pm$ 0.084 & 53.97 $\pm$ 5.52 & -242.21 & -1.521 $\pm$ 0.109 & -2.81 $\pm$ 2.29 & 51.41 $\pm$ 4.89 & -250.67 & CPL \\ 
{[52.66 , 55.15]} & -1.529 $\pm$ 0.030 & 132.77 $\pm$ 5.37 & -262.53 & -1.527 $\pm$ 0.030 & -8.32 $\pm$ 1.95 & 133.19 $\pm$ 5.47 & -271.82 & CPL \\ 
{[55.15 , 59.44]} & -1.697 $\pm$ 0.045 & 51.68 $\pm$ 3.60 & -310.38 & -1.695 $\pm$ 0.051 & -9.11 $\pm$ 1.92 & 52.17 $\pm$ 3.61 & -316.46 & CPL \\ 
{[59.44 , 62.58]} & -1.497 $\pm$ 0.034 & 117.84 $\pm$ 4.67 & -292.39 & -1.497 $\pm$ 0.034 & -8.05 $\pm$ 1.85 & 117.75 $\pm$ 4.71 & -303.78 & CPL \\ 
{[62.58 , 65.20]} & -1.772 $\pm$ 0.088 & 38.41 $\pm$ 12.19 & -267.84 & -1.766 $\pm$ 0.091 & -3.47 $\pm$ 2.09 & 39.29 $\pm$ 9.84 & -272.76 & CPL \\ 
{[65.20 , 68.02]} & -1.597 $\pm$ 0.041 & 166.11 $\pm$ 14.93 & -271.10 & -1.601 $\pm$ 0.040 & -6.06 $\pm$ 2.03 & 167.10 $\pm$ 14.28 & -284.43 & CPL \\ 
{[68.02 , 72.50]} & -1.539 $\pm$ 0.043 & 127.30 $\pm$ 7.95 & -298.35 & -1.543 $\pm$ 0.044 & -7.68 $\pm$ 2.01 & 126.77 $\pm$ 7.99 & -310.06 & CPL \\ 
{[79.47 , 84.14]} & -1.590 $\pm$ 0.068 & 99.18 $\pm$ 10.05 & -306.67 & -1.591 $\pm$ 0.076 & -5.57 $\pm$ 2.12 & 98.78 $\pm$ 9.75 & -322.78 & CPL \\ 
\end{longtable}

\end{document}